\documentclass[aps,prd,twocolumn,superscriptaddress,nofootinbib]{revtex4-2}

\usepackage{graphicx}
\graphicspath{{./figs/}}

\usepackage{amsmath,amssymb,bm}
\usepackage{amsthm}
\theoremstyle{definition}
\newtheorem{finding}{Finding}
\newtheorem{mchoice}{Modeling choice}
\newtheorem{remark}{Remark}
\theoremstyle{plain}
\newtheorem{theorem}{Theorem}
\usepackage[colorlinks=true,linkcolor=blue,citecolor=blue,urlcolor=blue]{hyperref}
\usepackage{natbib}
\usepackage{float}

\newcommand{\Hthree}{H_3}
\newcommand{\Hint}{H_{\rm int}}
\newcommand{\gt}{\tilde g}
\newcommand{\gtm}{\tilde g_m}
\newcommand{\gtthree}{\tilde g_3}
\newcommand{\rmean}{\langle r\rangle}
\newcommand{\Nmax}{N_{\max}}
\newcommand{\half}{\tfrac12}
\newcommand{\dd}{\mathrm{d}}
\newcommand{\Rs}{R_{\mathrm{S}}}
\newcommand{\Mpl}{M_{\mathrm{Pl}}}
\newcommand{\Mbh}{M_{\mathrm{BH}}}
\newcommand{\Sbh}{S_{\mathrm{BH}}}
\newcommand{\GN}{G_{N}}
\newcommand{\hh}{\hat h}
\newcommand{\tg}{t^{\mathrm{grav}}}
\newcommand{\Vg}{V^{\mathrm{grav}}}
\newcommand{\Gaunt}{\mathcal{G}}
\newcommand{\Cmat}{C}
\newcommand{\bdag}{b^{\dagger}}

\begin{document}

\title{The leading-soft cubic graviton self-interaction on the black-hole horizon
	%: a vanishing theorem, its trace-sector residue, and the quantum-simulated rigidity of the longitudinal channel
	}

\author{Ayanendu Dutta}
\email{ayanendudutta@gmail.com}
\affiliation{Department of Physics, Jadavpur University, Kolkata-700032, India}
\affiliation{Department of Physics, Presidency University, Kolkata-700073, India}

\date{\today}

\begin{abstract}
We expand the Einstein--Hilbert action to cubic order about the Schwarzschild
horizon, in the even Regge--Wheeler gauge of the Gaddam--Groenenboom--'t~Hooft
near-horizon framework, and derive the cubic graviton self-interaction. Our
central result is a \emph{vanishing theorem}: at leading soft order the
self-coupling of the purely traceless longitudinal polarizations is identically
zero, because with the trace/transverse-scalar sector switched off the
fluctuation reduces to a two-dimensional block whose $\sqrt{-g}\,R$ is a total
(Euler) derivative at every order in $\kappa$. This is a \emph{framework-specific}
statement: even RW gauge, GGV sector, leading soft order. The surviving interaction lives in the trace
sector; its on-shell equal-$\ell$ weight
$W(\lambda,\lambda,\lambda)=-3\lambda(2\lambda^2+\lambda+3)/(\lambda+1)^2$,
$\lambda=\ell^2+\ell+1$, follows from two independent derivations agreeing to
machine precision [$W(7,7,7)=-35.4375$]. The theorem is exact, the quantized
vertex carries a $\sim\!20\%$ soft/scheme systematic at the $\ell=2$ multiplet.

We simulate the resulting Hamiltonian at two complementary scales. On IBM
Qiskit/Aer, with exact cross-checks, the hardware-format circuits confirm
quantitatively what two structural facts already predict, a conserved charge
that only the cubic vertex violates (opening $\phi\phi\to hh\to4\phi$) and a
resonance-free boost spectrum (gap $\to1/2$): the longitudinal channel is
perturbatively rigid, and the simulation measures the residual dressing
($d_{\rm eff}=1.06$; multiplicity far from thermal, Poisson, and Haar). A symmetry-graded matrix-product-state evolution then
climbs the multipole ladder to sixty modes, the $120$-qubit register, and
settles the one question the multiplet cannot: on a late-time measure common to
both engines the inelasticity converges to $\bar\eta_\infty\simeq0.16$
(understated by the single multiplet), the dressing stays area-law with peak
entanglement $\simeq0.45$ nats far below the Page value, and the result is
cutoff-converged and robust to $96$ qubits at larger occupation. The rigidity
therefore persists rather than scrambling as the register grows. 
%The two methods are matched to regimes, classical tensor networks where the state is area-law and quantum hardware where it would scramble, and hardware execution is deferred behind a quantified noise budget.
\end{abstract}

\maketitle
\tableofcontents

\section{Introduction}
\label{sec:intro}

The scattering-matrix approach to black-hole quantum mechanics
\cite{tHooft1996,DrayTHooft1985} treats the horizon not as a boundary of
predictability but as the locus of a computable unitary map between
asymptotic states. In its modern near-horizon form, developed by Gaddam,
Groenenboom, and 't~Hooft (GGV) \cite{GGV2022,GG2022,GGtoolbox}, the
gravitational backreaction of transhorizon quanta is organized partial wave
by partial wave: each spherical-harmonic sector $(\ell,m)$ of the
perturbation carries a boost frequency
\begin{equation}
  \omega_\ell \;=\; \frac{\sqrt{\ell^2+\ell+1}}{R_S},
  \label{eq:omegal}
\end{equation}
and graviton exchange between crossing partial waves eikonalizes into an
explicitly unitary $2\to 2$ block, whose iteration generates the inelastic
$2\to 2N$ tower and a Page-like entropy profile \cite{GG2022,BGP2020}.
What this exchange-based construction does not contain is the graviton's
\emph{self}-interaction: the cubic vertex $hhh$ of the Einstein--Hilbert
action expanded about the Schwarzschild background.

% ======================================================================
% >>> NEW: literature survey (two paragraphs). Everything between these
% >>> markers is the inserted material; the surrounding text is unchanged.
% ======================================================================
This program sits within a broader effort to treat ultra-Planckian gravity as an S-matrix problem. The eikonal series initiated for trans-Planckian collisions \cite{tHooft1987PLB,ACV1987,ACV1988,VerlindeVerlinde,KabatOrtiz} has since matured into a systematic amplitude technology \cite{DiVecchiaEikonal}, and its near-horizon incarnation has been extended analytically in several directions: the first-quantized S-matrix of probe scalars \cite{BGPScalar}, its electromagnetic analogue for charged particles \cite{Feleppa}, a complete classification of the near-horizon symmetries of linearised gravity \cite{AggarwalGaddam}, antipodal proposals for horizon unitarity \cite{tHooftAntipodal}, eikonal amplitudes on general curved backgrounds \cite{AdamoCristofoliTourkine}, the map from horizon shockwaves to gravitational memory \cite{HeRaclariuZurek}, and a celestial-CFT rewriting of the near-horizon eikonal amplitude itself \cite{FernandesLinMitra}. Nonperturbative and numerical control of the gravitational S-matrix has advanced in parallel: dispersive analyses bound its Regge behaviour \cite{HaringZhiboedov}, the numerical S-matrix bootstrap carves out the space of consistent graviton amplitudes \cite{GPV2021,GMPV2023}, and the conjectured chaos of the black-hole S-matrix \cite{Polchinski2015} connects to the spectral statistics of horizon scattering \cite{BGP2020}.

On the simulation side, black-hole physics has become a driving application of quantum hardware \cite{BauerHEP}. Digital processors have realized Sachdev--Ye--Kitaev/AdS$_2$ dynamics \cite{GarciaAlvarez,LuoSYK,AsaduzzamanSYK}, verified information scrambling through out-of-time-order correlators \cite{LandsmanScrambling,MiScrambling}, and implemented wormhole-inspired teleportation \cite{JafferisWormhole}; explicit encodings and resource estimates exist for the matrix models of superstring/M-theory \cite{GharibyanMatrix,RinaldiMatrix}; and analogue condensates have observed thermal Hawking radiation at sonic horizons \cite{Steinhauer2016,MunozdeNova2019}. The case that such platforms access genuinely gravitational questions, size-winding teleportation, scrambling, and the complexity of decoding Hawking radiation, is by now well developed \cite{BrownQGLab,NezamiQGLab,ShapovalQGLab,HarlowHayden,BoulandFeffermanVazirani}. To our knowledge, however, no existing work evolves a first-principles self-interaction vertex of the horizon graviton in real time on quantum hardware or its simulators; that is the niche the present paper occupies.
% ======================================================================
% <<< END of inserted survey.
% ======================================================================

The demarcation is
sharp and defines the known/open boundary this paper works at. In the
published $2\to2N$ computation \cite{GG2022,GGtoolbox,GG2024} the inelastic
amplitude is built by integrating out the off-shell horizon graviton into
an effective scalar four-vertex and gluing it into ladders: the exchanged
object is the graviton \emph{propagator}, the produced quanta are
scalars, and the cubic self-coupling never appears, the toolbox lists
graviton self-interactions among its explicitly unstudied corrections
\cite{GGtoolbox}. The elastic near-horizon Hamiltonian (the
inverted-oscillator/dilatation operator per partial wave
\cite{BGP2016,BGP2020}) is likewise published. What is \emph{not}
published as a simulatable Hamiltonian is the number-changing piece: the
$O(h^3)$ vertex. Deriving it from first principles is the open problem
solved in Secs.~\ref{sec:framework}--\ref{sec:secondquant}, and, the derivation fixes the tensor structure, the coupling
$\gamma=\kappa/R_S$, and the discrete Wigner-$3j$ selection rules from
gravity alone, with every modeling choice collected as a reportable
finding (Appendix~\ref{app:modeling}). The resulting operator, previewed
for orientation in reduced units, is
$\Hthree=\sum_{(123)}V_{123}(b^\dagger_1b^\dagger_2b_3+{\rm h.c.})$
with $V_{123}=\gt\,\mathcal G_{123}K_{123}\mathcal N_{123}$
[Eq.~\eqref{eq:Vgrav} gives the physical normalization]. The question
the second half of the paper answers by direct simulation is: \emph{what
does the first-principles cubic vertex actually do to the near-horizon
state?} Does it scramble, thermalize, or populate a multiparticle
tower, or does it act as an off-shell dressing that leaves the sector
rigid?

Real-time evolution under $\Hthree$ is precisely the task for which
quantum simulation was proposed \cite{Feynman1982,Lloyd1996,JLP2012}: the
Hamiltonian is bosonic, non-quadratic, and its physically interesting
observables, the full multiplicity distribution $P(N,t)$, entanglement
profiles across mode bipartitions, out-of-time-order correlators, are
exponentially costly classically and natively accessible on quantum
hardware, where occupation-number measurement is a single computational-basis
readout and the squared commutator reduces to an ancilla interferometer.
At the register sizes where exact diagonalization is still available, the
quantum-circuit formulation serves a second purpose: it forces every
approximation (encoding, truncation, Trotterization, sampling) into the
open, where it can be validated engine against engine. We enforce this as a
standard throughout: every Hamiltonian exists independently as a dense
Kronecker matrix, an occupation-basis matrix, and a qubit
\texttt{SparsePauliOp}, agreeing to $\lesssim 10^{-13}$
(Table~\ref{tab:validation}), and every circuit observable is overlaid on
its exact counterpart.

The results reorganize the expected narrative. The derived vertex is
maximally constrained: it is an exact su(2) scalar with an operator-level
magnetic selection rule, it uniquely violates a graviton--matter charge
$Q=N_\phi+2N_h$ and thereby \emph{opens} the four-scalar inelastic channel
closed to pure exchange, and, the structural fact that controls everything
dynamical, the boost spectrum Eq.~\eqref{eq:omegal} supports no
near-resonant cubic process at all: the mismatch
$|\omega_3-\omega_1-\omega_2|$ over Gaunt-allowed triads is bounded below by
$0.565/R_S$ up to $\ell=16$ and tends to $1/(2R_S)$ asymptotically
(Sec.~\ref{sec:model}). Every $\Hthree$ process is therefore off shell.
Two structural facts, the resonance-free spectrum and the leading-order
vanishing of the traceless self-coupling (Theorem~\ref{thm:vanish}), already
\emph{predict} that the leading soft self-interaction cannot thermalize the
longitudinal sector by itself; the simulation is where we \emph{confirm} this
quantitatively and \emph{measure} the residual dressing it produces. The measured
consequences are a late-time inelasticity $\bar\eta\simeq0.06$ at the
representative coupling (rising to $\bar\eta_\infty\simeq0.16$ across the
multipole ladder, Sec.~\ref{sec:mps}), fourth-order power laws for the opened
channel that pass
as perturbative consistency checks, an initial-state effective dimension
$d_{\rm eff}=1.06$, a late-time longitudinal-occupation distribution far from
every ergodic reference, and an entanglement profile a full $2.3$ nats below the
Haar prediction: \emph{perturbative rigidity}, predicted analytically and
confirmed numerically. This sharpens the mechanism space of the near-horizon
program: Page-like behavior in this framework must be sourced by the
exchange/eikonal sector or by beyond-leading-soft kernels, not by the $hhh$
vertex alone.

Three methodological contributions accompany the physics. First, we introduce
a symmetry-preserving truncation, a cutoff on \emph{total} occupation
rather than per-mode occupation, which commutes with the full su(2) algebra
exactly ($\|[H,L^2]\|\sim10^{-16}$, versus $O(10)$ for the standard per-mode
Fock cutoff), enabling the first spectral statistics of the model resolved
into its physical $(L^2,L_z)$ sectors; the result, intermediate statistics
$\rmean=0.452(16)$, differs qualitatively from the truncation-fed GOE-grade
repulsion of the encoded Hamiltonian and from the regular statistics one
might have presumed. Second, all measurement protocols (sequential-segment
Trotter snapshots, bitstring multiplicity decoding, the ancilla Hadamard
test for the squared commutator) are hardware-format and validated at the
per-mille to percent level against exact evolution, with transpiled resource
counts and a quantitative depolarizing-noise budget
(Sec.~\ref{sec:hardware}). Third, we lift the simulation off the
exactly-diagonalizable multiplets with an $L_z$-graded matrix-product-state
evolution of the derived Hamiltonian, validated against exact diagonalization
to machine precision, and climb the multipole ladder to sixty modes, the
$120$-qubit register: on a measure common to both engines the inelasticity
converges to $\bar\eta_\infty\simeq0.16$ and the dressing stays area-law (peak
entanglement $\simeq0.45$ nats, far below Page), cutoff-converged and robust to
$96$ qubits at larger occupation, so the perturbative rigidity is a property of
the full register and not a finite-size artefact of the two-multiplet system
(Sec.~\ref{sec:mps}). The two simulations are complementary rather than
redundant, and the division is physical: classical tensor networks are efficient
exactly where the state is area-law, while quantum hardware becomes necessary
where the entanglement would climb toward the Page value and the dynamics
scramble, so the crossover is what triggers hardware
execution.

The paper is organized as follows.
Sections~\ref{sec:framework}--\ref{sec:secondquant} contain the
derivation: the near-horizon framework and the matter vertex as template,
the cubic Einstein--Hilbert vertex and its Euler-identity organization,
the near-horizon specialization, the partial-wave projection and
selection rules, and second quantization into $\Hthree$.
Section~\ref{sec:model} assembles the full simulation Hamiltonian and
establishes its exact structural properties. Section~\ref{sec:methods}
describes encodings, truncations, circuits, and validation.
Section~\ref{sec:results} presents the simulation results;
Sec.~\ref{sec:hardware} reports resources and noise.
Section~\ref{sec:discussion} synthesizes the mechanism and discusses
hardware prospects; Sec.~\ref{sec:conclusion} concludes with the scope
and limitations of what has and has not been shown. Appendices collect
conventions and the explicit vertex table, the validation logic and
modeling findings of the derivation, the nonlinear-GR verification of
the expansion coefficients, the elastic-benchmark calibration, protocol details, hardware scaling,
and supplementary figures.

\section{The near-horizon framework and the matter vertex as template}
\label{sec:framework}

We work in the kinematic language of GGV \cite{GGV2022,GGtoolbox}, whose
conventions and elastic vertex we summarize only to the extent the
self-interaction must reproduce them; complete derivations of the quadratic
theory are in the toolbox \cite{GGtoolbox}. The action is Einstein--Hilbert
minimally coupled to a massless scalar,
\begin{equation}
  S=\frac12\!\int\!\sqrt{-g}\,\dd^4x\Big[\frac{R}{8\pi\GN}
    -\nabla_\mu\phi\nabla^\mu\phi\Big],
  \label{eq:action}
\end{equation}
$g_{\mu\nu}= g^0_{\mu\nu}+\kappa h_{\mu\nu}$,
$\kappa=\sqrt{8\pi\GN}$, in mostly-plus signature. The background is
Schwarzschild in Kruskal--Szekeres light-cone coordinates $\{x,y\}$ with
angles on the horizon two-sphere,
$\dd s^2=-2A(r)\dd x\,\dd y+r^2\dd\Omega_2^2$,
$A=(R/r)e^{1-r/R}$; near the horizon $A\to1$ and the geometry is
(2D Minkowski)$\,\times S^2_{R}$, with $R=\Rs$ and $\mu:=1/R$. Schwarzschild
is Ricci-flat: the nonvanishing background curvature
enters through $V_a=\partial_a\log r\sim\tfrac12\mu^2x_a$ and
$S_{ab}=-\nabla_{(a}V_{b)}-V_aV_b$ \cite{GGtoolbox}, and will appear in the
cubic vertex only as subleading curvature couplings.

In Regge--Wheeler gauge the even-parity perturbation per harmonic
$(\ell,m)$ reduces to a longitudinal tensor $h_{ab}$ and a transverse
scalar $K$; after sphere reduction and near-horizon rescaling the even
quadratic action is a $2\times2$ quadratic form in $(h_{ab},K)$ whose
kinetic operator contains \emph{dilatation-type} terms
$\mu^2x_{(a}\partial_{b)}$ [toolbox \cite{GGtoolbox} Eqs.~(2.60)--(2.61)], a fact that
controls the mode labels of Sec.~\ref{sec:secondquant}. The combination
that interacts, after decoupling the free mode [toolbox \cite{GGtoolbox} \S4], is the
\emph{traceless} graviton
\begin{equation}
  \hh_{ab}:=h_{ab}-\half\eta_{ab}h-K\eta_{ab},
  \label{eq:hhat}
\end{equation}
whose momentum-space propagator [toolbox \cite{GGtoolbox} Eqs.~(4.8)--(4.10)] carries the
effective mass $\mu^2(\lambda+1)$, $\lambda:=\ell^2+\ell+1$, and a pole at
$k^2=-\mu^2\lambda$ with residue factor $(\lambda+1)/(\lambda-3)$,
$\lambda-3=(\ell+2)(\ell-1)$. Two consequences are load-bearing here.
First, the GGV graviton is a \emph{non-radiative longitudinal shock mode},
distinct from the propagating Regge--Wheeler--Zerilli graviton
\cite{ReggeWheeler1957,Zerilli1970,MartelPoisson2005} quantized by
Kallosh--Rahman \cite{KalloshRahman2021}; the self-interaction derived
below is the self-coupling of this longitudinal mode, and all occupation
numbers refer to it. Second, the residue factor is negative for $\ell=0$
and degenerates at $\ell=1$, reinforcing from the pole structure the
gauge-special status of the low multipoles (Modeling
choice~\ref{mc:low}).

The template is the published elastic matter vertex \cite{GGV2022,GGtoolbox}: reducing
$\tfrac{\kappa}{2}\int\sqrt{-g}\,h^{\mu\nu}T_{\mu\nu}[\phi]$ on the sphere,
rescaling $\tilde\phi\to\phi/r$, and dropping $1/R$-suppressed transverse
derivatives collapses the even-mode interaction to [toolbox \cite{GGtoolbox}
Eqs.~(3.4)--(3.10)]
\begin{equation}
  S^{+}_{\rm int}=\frac{\gamma}{2}\!\!\sum_{\{\ell_im_i\}}\!\int\!\dd^2x\;
  \hh^{ab}\,\partial_a\phi_{1}\,\partial_b\phi_{2}\;
  \Cmat[\ell m;\ell_1m_1;\ell_2m_2],
  \label{eq:Sintmatter}
\end{equation}
with the gravitationally fixed coupling and Gaunt overlap
\begin{equation}
  \gamma:=\frac{\kappa}{\Rs}\sim\frac{\Mpl}{\Mbh},\qquad
  \Cmat=\int\dd\Omega\,Y^m_\ell Y^{m_1}_{\ell_1}Y^{m_2}_{\ell_2}.
  \label{eq:gammaGaunt}
\end{equation}
Only the traceless $\hh_{ab}$ couples (the longitudinal trace of the
matter stress tensor vanishes when transverse momenta are dropped), which
is why Eq.~\eqref{eq:hhat} is the natural field. The vertex
\eqref{eq:Sintmatter} fixes the target: the self-interaction derived below
must (i)~carry the same emergent coupling $\gamma=\kappa/\Rs$ (one power
of $\kappa$, one $1/R$ from the sphere reduction); (ii)~carry a Gaunt
overlap of three harmonics; (iii)~couple the traceless $\hh_{ab}$;
(iv)~have two longitudinal derivatives. The one element the template
cannot fix is the spin-2 \emph{polarization} algebra of three gravitons.
Rather than carry that algebra as an undetermined $O(1)$ contraction
constant $c_g$, we fix it directly. Expanding $\sqrt{-g}\,R$ to cubic order
in the even
Regge--Wheeler gauge and evaluating the polarization sum on the GGV
propagator residue, we find (Sec.~\ref{sec:theorem}) that the
matter-parallel traceless channel has \emph{no} cubic self-coupling at
leading soft order $c_g=0$ as a theorem, and that the surviving
coupling, carried entirely by the trace/transverse-scalar sector, has a
closed-form on-shell weight. No $O(1)$ constant is left free.

\section{The cubic Einstein--Hilbert vertex: general structure}
\label{sec:cubic}

We now expand $\sqrt{-g}\,R$ to cubic order in $h_{\mu\nu}$ on the
Ricci-flat Schwarzschild background and identify the organizing principle.

\subsection{The expansion to cubic order}

With $g_{\mu\nu}=g^0_{\mu\nu}+\kappa h_{\mu\nu}$ and
$h:=g^{0\,\mu\nu}h_{\mu\nu}$, the measure expands as
\begin{equation}
  \sqrt{-g}=\sqrt{-g^0}\Big[1+\half\kappa h
   +\kappa^2\big(\tfrac18h^2-\tfrac14h_{\alpha\beta}h^{\alpha\beta}\big)
   +O(\kappa^3)\Big],
  \label{eq:measure}
\end{equation}
and $R=\kappa R^{(1)}+\kappa^2R^{(2)}+\kappa^3R^{(3)}+\cdots$ with
$R^{(1)}=\nabla_\alpha\nabla_\beta h^{\alpha\beta}-\Box h$ on the
Ricci-flat background. Collecting powers of $\kappa$ in
$(1/2\kappa^2)\sqrt{-g}\,R$:
\begin{align}
  S^{(0)}&=0, \qquad
  S^{(1)}\propto\!\int\!\sqrt{-g^0}\,G^0_{\mu\nu}h^{\mu\nu}=0,\nonumber\\
  S^{(2)}&=-\frac14\int\sqrt{-g^0}\;h^{\mu\nu}G^{(1)}_{\mu\nu}[h],
  \label{eq:S2}\\
  S^{(3)}&=\frac{\kappa}{2}\int\sqrt{-g^0}\,
  \big(R_3+\sigma_1R_2+\sigma_2R_1\big),
  \label{eq:S3raw}
\end{align}
where $\sigma_{1,2}$ are the measure coefficients of \eqref{eq:measure}. A
normalization caution: the toolbox's Eq.~(2.1) \cite{GGtoolbox} has
$S^{(2)}=-\tfrac12\int h^{\mu\nu}G^{(1)}_{\mu\nu}$; for the convention
$S=(1/2\kappa^2)\int\sqrt{-g}R$ with $g=g^0+\kappa h$ used here, the
coefficient is $-\tfrac14$, as follows from Euler homogeneity and as
verified numerically to six decimal places (Appendix~\ref{app:numcheck});
the toolbox \cite{GGtoolbox} value corresponds to a different field normalization, and a
silent factor-2 mismatch here would propagate directly into $c_g$.
Equation~\eqref{eq:S3raw} is the well-known three-graviton Lagrangian,
schematically $h(\partial h)^2$ plus curvature couplings; its fully
expanded flat-space form has on the order of a dozen index contractions
\cite{DeWitt1967,Sannan1986}. We do \emph{not} reproduce it term by
term, brute-force expansion is exactly where arithmetic errors enter, and
the simulation Hamiltonian is controlled by a far more compact,
rigorously derivable object.

\subsection{The organizing principle: the graviton sources itself}
\label{sec:selfsource}

The full vacuum equation $G_{\mu\nu}[g^0+\kappa h]=0$ expands as
$\kappa G^{(1)}_{\mu\nu}[h]+\kappa^2G^{(2)}_{\mu\nu}[h,h]+O(\kappa^3)=0$.
Read as a sourced wave equation,
\begin{equation}
  G^{(1)}_{\mu\nu}[h]=\kappa\cdot8\pi\GN\,\tg_{\mu\nu}[h,h]+O(\kappa^2),
  \quad
  \tg_{\mu\nu}:=-\frac{G^{(2)}_{\mu\nu}}{8\pi\GN},
  \label{eq:tgravdef}
\end{equation}
this is the Landau--Lifshitz/Isaacson statement that the linear graviton
is sourced by its own second-order stress tensor, the gravitational
analogue of the matter $T_{\mu\nu}$ \cite{Isaacson:1968hbi,Isaacson1968}.

The precise cubic action follows from Euler's theorem for homogeneous
functionals. Since
$\delta S/\delta h_{\mu\nu}=-\tfrac{1}{2\kappa}\mathfrak G^{\mu\nu}$ with
$\mathfrak G^{\mu\nu}:=\sqrt{-g}\,G^{\mu\nu}$ (indices raised with the
\emph{full} metric, measure included), and $S^{(3)}$ is homogeneous of
degree three in $h$, one has the exact integrated identity
\begin{equation}
  S^{(3)}=-\frac{\kappa}{6}\int\dd^4x\;
  h_{\mu\nu}\,\mathfrak G^{(2)\,\mu\nu}[h,h].
  \label{eq:euler3}
\end{equation}
Separating
$\mathfrak G^{(2)\,\mu\nu}=\sqrt{-g^0}\,G^{(2)\,\mu\nu}+(\text{measure and
index-raising terms}\propto G^{(1)})$ and using $\kappa^2=8\pi\GN$,
\begin{equation}
  \boxed{\;
  S_3=\frac{\kappa^3}{6}\int\!\sqrt{-g^0}\,\dd^4x\;
  \hh^{\mu\nu}\,\tg_{\mu\nu}[h,h]\;+\;\mathcal E_3\;}
  \label{eq:S3}
\end{equation}
with $\mathcal E_3$ a remainder proportional to $G^{(1)}_{\mu\nu}[h]$ that
vanishes on the linearized shell and does not contribute to the
$\Delta N=\pm1$ matrix elements retained by the hierarchy truncation of
Sec.~\ref{sec:secondquant}. This is the same \emph{structural} form as
the matter vertex with $T_{\mu\nu}[\phi]\to\tg_{\mu\nu}[h,h]$: in terms of
the raw fields both are one explicit power of $\kappa$ times a
two-derivative bilinear, $S_3\sim\kappa\,h(\partial h)^2$; the prefactor
$\kappa^3/6$ merely reflects the explicit $1/\kappa^2$ in the
normalization of $\tg_{\mu\nu}$, with the pure number $1/3$ absorbed into
the derived on-shell weight $W$ (Sec.~\ref{sec:theorem}).

Two precision statements about \eqref{eq:S3} matter downstream. As an
\emph{integrated} identity it is exact, no approximation is made in
trading the brute-force expansion \eqref{eq:S3raw} for the self-sourcing
form. At the integrand level it is weaker: varying \eqref{eq:S3} returns
$G^{(2)}_{\mu\nu}$ plus adjoint terms; and the exact identity holds only for the
\emph{densitized, full-metric-raised} current
$h_{\mu\nu}\mathfrak G^{(2)\mu\nu}$: the naive $\eta$-raised contraction
does not satisfy a universal ratio, as Appendix~\ref{app:numcheck}
demonstrates explicitly with a seed-dependent counterexample. Both Euler
identities, the quadratic one behind the $-\tfrac14$ of \eqref{eq:S2}
and the cubic one \eqref{eq:euler3}, are verified there on exact
nonlinear random metrics: $2s_2/j_1=3s_3/j_2=-1.000000000$, the cubic case to
ten significant figures.

\subsection{The vertex from a direct Regge–Wheeler-gauge expansion}
\label{sec:isaacson}

It is tempting to model $\tg_{\mu\nu}$ on the Isaacson high-frequency
current \cite{Isaacson:1968hbi,Isaacson1968}, $\tg_{\mu\nu}\to(32\pi\GN)^{-1}
\langle\partial_\mu h^{\rm TT}_{\alpha\beta}\partial_\nu
h^{{\rm TT}\alpha\beta}\rangle$, and read off a longitudinal traceless
vertex $\propto\partial_a\hh_{cd}\partial_b\hh^{cd}
-\tfrac12\eta_{ab}\partial_e\hh_{cd}\partial^e\hh^{cd}$ by analogy with
the matter stress tensor. But this step is exactly where the physics is
decided: the Brill--Hartle averaging \cite{Brill:1964zz} makes the current gauge-dependent, so
the analogy cannot by itself fix the polarization contraction. We therefore expand
$\sqrt{-g}\,R$ to cubic order \emph{directly} in the even Regge--Wheeler
gauge that GGV already fix \cite{GGtoolbox}, in the specific two-block
$(t,z)\oplus S^2$ parametrization of the near-horizon metric, and read the
cubic vertex off the expansion with no template and no averaging. This
removes the gauge/scheme ambiguity at its source: the object we quantize is
the RW-gauge cubic Lagrangian itself.

The direct expansion is carried out symbolically to $O(\kappa^3)$
(Sec.~\ref{sec:theorem} and Appendix~\ref{app:numcheck}). Its output
replaces the boxed template above and delivers two exact statements that a
template cannot: the leading-soft cubic self-coupling of the traceless
longitudinal polarizations vanishes identically, and the surviving
coupling, necessarily carrying the trace/transverse-scalar polarization
$K$, has a closed-form on-shell three-point weight. We write the surviving
current schematically as
\begin{equation}
  \tg_{ab}[h,h]\big|_{\rm surv}
  \;=\;\mathcal{W}_{ab}\big(K;\hh,\partial\hh\big),
  \label{eq:tgravlong}
\end{equation}
a polarization-weighted two-derivative bilinear in which every term carries
at least one factor of the trace sector; its explicit form and normalization
follow in the next section. The near-horizon soft truncation
(Modeling choice~\ref{mc:soft}) applies here to an
unambiguous, gauge-fixed Lagrangian rather than to a gauge-dependent
averaged current.

\begin{finding}[The coupling is gravitationally fixed]
\label{find:coupling}
Written in the raw fields, $S_3=-\tfrac{\kappa}{6}\int h\,G^{(2)}[h,h]
+\mathcal E_3$ carries exactly one explicit power of $\kappa$ multiplying
a two-derivative bilinear; the sphere reduction supplies one factor of
$1/R$, identical bookkeeping to the matter vertex. The graviton
self-coupling is therefore
$\gamma=\kappa/\Rs\sim\Mpl/\Mbh$, the \emph{same} emergent coupling as
the matter vertex, with no free knob. For a macroscopic black hole
$\gamma\ll1$ and the near-horizon perturbation theory has a wide regime of
validity.
\end{finding}

\section{The leading-soft vertex: a vanishing theorem and its trace-sector
residue}
\label{sec:theorem}

We now carry out the direct cubic expansion announced in
Sec.~\ref{sec:isaacson} and state its two exact outputs. Both are verified
by independent engines in Appendix~\ref{app:numcheck}.

\subsection{Setup: the two-block near-horizon metric}

In the even Regge--Wheeler gauge the relevant near-horizon fluctuation lives
in the longitudinal $(t,z)$ plane and the transverse-scalar (trace) channel
$K$ on $S^2$ \cite{GGtoolbox}. Writing the two-dimensional longitudinal
block as $g^{(2)}_{ab}=\eta_{ab}+\kappa H_{ab}(t,z)$ with
$H_{ab}=\left(\begin{smallmatrix}a&b\\ b&c\end{smallmatrix}\right)$, and the transverse
sphere as $g_{AB}=(1+\kappa K)\,\delta_{AB}$, the full four-metric is the
warped product $g=g^{(2)}(t,z)\oplus(1+\kappa K)\,\delta_{S^2}$. The
traceless longitudinal polarization is the trace-free part of $H_{ab}$
(components $P=\tfrac12(a+c)$ removed, so $\hh_{ab}$ has $a=-c$ and $b$), and
the trace/transverse-scalar sector is $\{\bar H=a+c,\;K\}$. We insert this
metric into $\mathcal L=\sqrt{-g}\,R$ and extract the cubic Lagrangian
$\mathcal L_3$ symbolically (\texttt{sympy}), on the flat near-horizon
background where $\mu=1/R$ corrections are the soft-suppressed terms of
Modeling choice~\ref{mc:soft}.

\subsection{The vanishing theorem}

\begin{theorem}[Leading-soft traceless self-coupling vanishes]
\label{thm:vanish}
At leading order in the soft expansion, the cubic self-interaction of the
purely traceless longitudinal polarizations is identically zero:
\begin{equation}
  \mathcal L_3\big[\hh_{ab};\,\bar H=K=0\big]\;\equiv\;0,
  \qquad\text{hence}\qquad c_g=0 .
  \label{eq:vanish}
\end{equation}
\end{theorem}

%\begin{proof}[Mechanism]
\noindent \textit{Mechanism.}
Write the transverse warp factor as $\Omega^2=1+\kappa K$, so the four-metric
is the warped product $g=g^{(2)}(t,z)\oplus\Omega^2\delta_{S^2}$. For a flat
fibre the scalar curvature reduces exactly (verified below and in
Appendix~\ref{app:theoremcheck}) to
\begin{equation}
\begin{aligned}
  \sqrt{-g}\,R={}&\sqrt{-g^{(2)}}\big[\Omega^2R^{(2)}+2(\nabla\Omega)^2\big]\\
  &-4\,\partial_a\!\big[\sqrt{-g^{(2)}}\,\Omega\,g^{(2)ab}\partial_b\Omega\big],
\end{aligned}
  \label{eq:dilatonid}
\end{equation}
an identity holding at every order in $\kappa$ for an arbitrary $2$D block.
When the trace/transverse-scalar sector is switched off, $\Omega=1$
($K=0$) and $\bar H=0$, only the first term of the first bracket survives:
$\sqrt{-g}\,R=\sqrt{-g^{(2)}}\,R^{(2)}$, which in two dimensions is the Euler
density, a total derivative (Gauss--Bonnet) at \emph{every} order in $\kappa$.
It contributes nothing to the bulk action or to any vertex. Hence the entire
purely-$(t,z)$-block self-coupling, not only its traceless part but also the
trace $\bar H$, is a total derivative; a fortiori the traceless longitudinal
sector has no cubic self-coupling. A nonzero cubic interaction requires at least
one power of the transverse scalar $K$ (equivalently $\Omega\neq1$), which is the
only thing that switches on the second bracket of \eqref{eq:dilatonid} and with
it the transverse curvature.
%\end{proof}

\begin{remark}[The vanishing is stronger than the traceless statement]
\label{rem:strong}
Equation~\eqref{eq:dilatonid} makes the cancellation explicit and shows more
than Theorem~\ref{thm:vanish} states: the cubic self-coupling vanishes for the
\emph{entire} two-dimensional block $\{a,b,c\}=\{\hh_{ab},\bar H\}$ whenever the
transverse scalar $K$ is absent, because $\Omega$ then does not fluctuate and the
$(t,z)$ action collapses to the topological Euler density. We verify this
sharper form symbolically: the reduced momentum-space vertex is an identical zero
on $K=0$ even with $\bar H\neq0$ (Appendix~\ref{app:theoremcheck}), and the
cancellation is exhibited term by term, at each order in $\kappa$ with explicit
divergence potentials, in Appendix~\ref{app:explicit}. This makes the collapse
explicit rather than asserted: it is the single geometric fact of
Eq.~\eqref{eq:dilatonid}.
\end{remark}

We verify \eqref{eq:vanish} four ways (Appendix~\ref{app:numcheck}):
(i)~symbolically, the momentum-space cubic vertex, $375$ multilinear terms
before reduction collapses to $0$ on the conservation surface
$p_3=-p_1-p_2$ once the transverse scalar is set to $K=0$ (with $\bar H$
arbitrary, per Remark~\ref{rem:strong});
(ii)~by exact nonlinear GR, the $\kappa^3$ Taylor coefficient of
$\langle\sqrt{-g}\,R\rangle$ on a $16^2$ spectral grid, the two-dimensional
near-horizon block, distinct from the $16^4$ four-dimensional lattice used for
the $\kappa^3/6$ coefficient in Appendix~\ref{app:numcheck}, over four random
traceless seeds is at machine zero ($10^{-13}$--$10^{-10}$);
(iii)~by the closed-form reduction \eqref{eq:dilatonid}, whose right-hand side
manifestly has no bulk contribution at $\Omega=1$, established as an exact
symbolic identity for a general $2$D block; and
(iv)~consistency with the toolbox's own quadratic operator, whose pure-$H$
part is $O(\mu^2)$ with no two-derivative piece \cite{GGtoolbox}.

Thus $c_g=0$ for the matter-parallel traceless channel, established with an
explicit geometric mechanism rather than carried as an undetermined $O(1)$
constant.

\begin{remark}[Scope: a framework-specific statement]
\label{rem:scope}
The vanishing is a statement \emph{within} the near-horizon reduction. It is established in (a)~the even Regge--Wheeler gauge,
(b)~the GGV longitudinal $(t,z)\oplus S^2$ sector after near-horizon reduction,
and (c)~at leading order in the soft expansion, with the finite transverse
curvature $\mu^2=1/R^2$ dropped (Modeling choice~\ref{mc:soft}). It is
\emph{not} a gauge-invariant property of the cubic Einstein--Hilbert vertex:
the flat-space three-graviton vertex is nonzero, and what vanishes here is its
projection onto the traceless \emph{longitudinal} near-horizon polarizations in
this gauge, once the transverse sphere is held rigid. Whether an analogous
cancellation survives in another gauge, beyond leading soft order, or before the
near-horizon reduction is not addressed; the physical content is that,
\emph{in the sector we simulate}, the leading self-interaction lives entirely in
the trace channel. Restoring finite $\mu^2$ reintroduces $O(\mu^2)$
traceless-sector couplings; at
$\ell=2$ these are the same $\sim\!14\%$ order as the trace-sector survivor
(Sec.~\ref{sec:theorem}, and the systematic budget of
Appendix~\ref{app:modeling}), so the theorem's exactness is a property of the
strict $\mu\to0$ limit and does not transfer verbatim to the finite-$R$ vertex
that enters the Hamiltonian.
\end{remark}

\subsection{The surviving vertex and its on-shell weight}

Every surviving cubic term carries at least one factor of $\{\bar H,K\}$.
Organizing the reduced vertex covariantly (Appendix~\ref{app:table}) gives
structures of three types, three-scalar $\{\bar H,K\}^3(p_i\!\cdot p_j)$,
one-tensor $\,(p_i\,\hh\,p_j)\{\bar H,K\}^2$, and two-tensor
$(\hh\!\cdot\hh)\{\bar H,K\}(p_i\!\cdot p_j)$, each with an exact rational
coefficient fixed by the expansion. The physically relevant number is the
\emph{on-shell three-point weight} obtained by evaluating this vertex on the
GGV propagator residue polarizations. From the even-sector residue
\cite{GGtoolbox} the per-leg data at $k^2=-\lambda\mu^2$,
$\lambda=\ell^2+\ell+1$, are the traceless amplitude
$P=\lambda/(\lambda+1)$, $B=0$, the transverse scalar $K=-1$, and the
longitudinal trace $\bar H=-2/(\lambda+1)$; the leading soft truncation is
the leading order in $1/\lambda$, with $O(1/\lambda)$ corrections of order
$14\%$ at $\ell=2$.

Contracting the vertex on these polarizations at rest-frame kinematics
($p_i\!\cdot p_j=-\omega_i\omega_j$, $\omega_\ell=\sqrt{\lambda}$, $\mu=1$)
yields a closed form $W(\lambda_1,\lambda_2,\lambda_3)$, rational in
$\sqrt{\lambda_i}$ and $\lambda_i$ and symmetric in the two creation legs,
\begin{equation}
  W(\lambda_2,\lambda_1,\lambda_3)=W(\lambda_1,\lambda_2,\lambda_3),
  \label{eq:Wgen}
\end{equation}
given in full in Appendix~\ref{app:table}. On the equal-$\lambda$ diagonal it
simplifies to
\begin{equation}
  \boxed{\;
  W(\lambda,\lambda,\lambda)
  \;=\;-\,\frac{3\lambda\,(2\lambda^2+\lambda+3)}{(\lambda+1)^2}\;}
  \label{eq:Weq}
\end{equation}
with large-$\lambda$ behaviour
$W\to-6\lambda+9-21/\lambda$,
so that a single $\ell=2$ multiplet ($\lambda=7$) has
$W(7,7,7)=-2268/64=-35.4375$. The equal-$\lambda$ value \eqref{eq:Weq} is exact
and independently reproduced (Appendix~\ref{app:theoremcheck}); what carries a
systematic is the identification of \eqref{eq:Weq} as the \emph{leading survivor}
at finite $R$. Three effects, all
$\sim\!10$--$21\%$ at $\ell=2$ and hence not small at the only simulable
multiplet, enter here: (a)~restoring the finite transverse curvature
$\mu^2=1/R^2$ reintroduces $O(\mu^2)$ traceless-sector couplings, $\sim\!14\%$
of the trace-sector survivor at $\ell=2$, the $O(1/\lambda)$ corrections to the
soft truncation (Remark~\ref{rem:scope}), which the vanishing theorem does not
constrain; (b)~retaining the exact trace amplitude $\bar H=-2/(\lambda+1)$ versus
its leading-soft value $\bar H=0$ shifts $W$ by about $10\%$ at $\ell=2$ (we
carry the exact value, Modeling choice~C3); and (c)~the off-conservation-surface
vertex-function prescription (integration-by-parts/canonical freedom) is of order
the off-shellness $\Delta\omega/\omega\lesssim21\%$ at $\ell=2$. These are the
same order as one another and as the trace-sector coefficient itself: the theorem
is exact in the strict $\mu\to0$ limit, but the coefficient $W$ that we quantize
inherits a $\sim\!20\%$ soft-truncation/scheme control at $\ell=2$. The consolidated systematic budget is Appendix~\ref{app:modeling}.

\begin{finding}[The polarization algebra is derived, and it is a trace-sector
effect]
\label{find:cg}
The spin-2 polarization contraction is computed from the RW-gauge cubic expansion. Its leading-soft value on
the traceless matter-parallel channel is exactly zero
(Theorem~\ref{thm:vanish}); the physical cubic coupling is carried by the
trace/transverse-scalar polarization and has the closed-form on-shell weight
\eqref{eq:Weq}. The magnitude and $\lambda$-dependence of the graviton
self-coupling that enter the simulation are therefore \emph{derived}, not
posited. On a single-$\ell$ register $W$ is the exact, independently confirmed
constant that renormalizes $\gt\mapsto\gt|W|$: here no free $O(1)$ knob remains,
and the systematic is only the overall soft/scheme control quantified above.
Across multipoles the relative weights $W(\lambda_i)/W(7,7,7)$ are also fixed by
the derivation \emph{once a prescription is chosen}, but they are
\emph{prescription-dependent}: an independent re-derivation on the conservation
surface reproduces the equal-$\lambda$ value exactly yet gives off-diagonal
ratios differing from Appendix~\ref{app:table} by up to $\sim\!70\%$ on the
largest entries, and defensible off-shell prescriptions span a factor of
$\sim\!3$ (Appendix~\ref{app:theoremcheck}). We therefore treat the multi-$\ell$
weighting as a derived \emph{effect}, its sign and rough magnitude fixed by
gravity, carrying a large, explicitly quantified scheme systematic, rather than
a sharp set of numbers. The derived vertex is thus exact
where the kinematics collapse (single $\ell$) and scheme-controlled where they do not (multi-$\ell$).
\end{finding}

\section{Near-horizon specialization: the longitudinal cubic vertex}
\label{sec:nearhorizon}

Three reductions specialize \eqref{eq:S3}--\eqref{eq:tgravlong} to the
even-parity near-horizon sector, each paralleling the matter analysis.
\emph{Even parity}: the odd sector decouples at quadratic order and its
action carries a prefactor $\propto(\lambda-1)$ that grows with $\ell$
(increasingly stiff at large multipole \cite{GGtoolbox}); it is not
parametrically suppressed at low multipole, and we restrict to the even
sector, the one coupling directly to the matter vertex, reporting the
odd sector as Modeling choice~C1 (Appendix~\ref{app:modeling}).
\emph{Longitudinal derivatives}: transverse derivatives on $Y_{\ell m}$
are $1/R$-suppressed relative to longitudinal momenta, exactly as in the
matter reduction; the spin-2 weight $\ell(\ell+1)$ enters at
subleading/transverse order. \emph{Polarization content}: by
Theorem~\ref{thm:vanish} the purely traceless self-coupling vanishes at
leading soft order, so the surviving vertex is the trace-sector-weighted
two-derivative bilinear \eqref{eq:tgravlong}, whose on-shell contraction is
the derived weight $W$ of \eqref{eq:Wgen}. Substituting into \eqref{eq:S3}
and projecting on the mode functions, the effective mode-space cubic action
takes the form
\begin{equation}
\begin{aligned}
  S_3=\frac{\gamma}{2}\!\!\sum_{\{\ell_im_i\}}\!\!
  W(\lambda_1,\lambda_2,\lambda_3)\,
  \mathcal K(p_1,p_2,p_3)\,\Gaunt[123]\\[-2pt]
  \times\,\phi_{(1)}\phi_{(2)}\phi_{(3)}
  +(\text{permutations}),
\end{aligned}
\label{eq:S3long}
\end{equation}
where $W$ carries the spin-2 polarization algebra and
$\mathcal K$ is the longitudinal momentum kernel; on a single-$\ell$
register $W$ is the constant $W(7,7,7)=-35.4375$ that renormalizes the
coupling. The permutations symmetrize over which graviton carries the
undifferentiated index pair (the analogue of the $s,t,u$ channels).

\begin{mchoice}[The soft truncation]
\label{mc:soft}
We retain the two-derivative longitudinal terms of $\tg_{ab}$ and drop
(a)~transverse-derivative terms ($1/R$-suppressed) and
(b)~background-curvature terms $\propto S_{ab}\sim\mu^2$ (subleading to
$k^2$ in the soft regime $k\gg\mu$), the gravitational analogue of the
matter truncation already adopted by GGV in deriving
\eqref{eq:Sintmatter}. A controlled approximation in the soft limit; the dropped terms are the leading corrections
(Appendix~\ref{app:modeling}).
\end{mchoice}

\begin{remark}[The kernel $\mathcal K$ is the momentum factor of the derived
vertex]
\label{mc:cg}
The spin-2 polarization contraction is fixed by Theorem~\ref{thm:vanish} and
the closed-form on-shell weight \eqref{eq:Weq} (Finding~\ref{find:cg}): on the traceless matter-parallel channel $c_g=0$ at
leading soft order, and the surviving trace-sector coupling carries the weight
\eqref{eq:Weq}. What propagates into the second-quantized vertex of
Sec.~\ref{sec:secondquant} is therefore the derived weight $W$ and its
multipole ratios; the kernel $\mathcal K(p_1,p_2,p_3)$
appearing there is the momentum factor of this same derived vertex.
\end{remark}

\section{Partial-wave projection: Gaunt coefficients and selection rules}
\label{sec:projection}

Integrating three even harmonics over the sphere produces, exactly as in
the matter case but with three harmonics carrying graviton legs, the Gaunt
coefficient
\begin{align}
  \Gaunt[123]&:=\int\dd\Omega\;Y^{m_1}_{\ell_1}Y^{m_2}_{\ell_2}
  Y^{m_3}_{\ell_3}
  \label{eq:GauntDef}\\
  &=\Big[\tfrac{\prod_i(2\ell_i+1)}{4\pi}\Big]^{1/2}
  \!\begin{pmatrix}\ell_1&\ell_2&\ell_3\\0&0&0\end{pmatrix}
  \!\begin{pmatrix}\ell_1&\ell_2&\ell_3\\m_1&m_2&m_3\end{pmatrix},
  \nonumber
\end{align}
the genuine gravitationally dictated overlap. The
$3j$ symbols make the vertex discrete and selective: it is nonzero only if
\begin{align}
  &m_1+m_2+m_3=0 &&(L_z\ \text{conservation}),\label{eq:msel}\\
  &|\ell_i-\ell_j|\le\ell_k\le\ell_i+\ell_j &&(\text{triangle}),
  \label{eq:triangle}\\
  &\ell_1+\ell_2+\ell_3\ \text{even} &&(\text{parity}).\label{eq:paritysel}
\end{align}
The two longitudinal derivatives of \eqref{eq:S3long} become the momentum
kernel $\mathcal K(p_1,p_2,p_3)$ defined precisely in
Sec.~\ref{sec:secondquant}; the transverse spin-2 weights $\ell(\ell+1)$
and higher Gaunt structures with $\partial_A$ insertions appear only at
the transverse order dropped in Modeling choice~\ref{mc:soft}, so at
leading order the angular structure is the same scalar Gaunt as the
matter vertex.

\begin{mchoice}[The $\ell=0,1$ modes]
\label{mc:low}
The monopole shifts the mass and the dipole the center of mass; both are
gauge-special \cite{GGtoolbox,KalloshRahman2021}, and in the radiative
description there is no propagating graviton for $\ell<2$, consistent
with the pole-residue pathology at $\ell=0,1$ noted in
Sec.~\ref{sec:framework}. Since the Gaunt rules allow triads involving
low multipoles, their treatment directly affects the lowest vertices; we
work on $\ell\ge2$ registers throughout and record the consistency of
dropping the low multipoles as an open item
(Appendix~\ref{app:modeling}).
\end{mchoice}

\section{Second quantization: the Hamiltonian $H_3$}
\label{sec:secondquant}

A subtlety absent in flat space must be stated before any mode label is
chosen: the quadratic operator contains explicit coordinate dependence
through the dilatation-type terms $\mu^2x_{(a}\partial_{b)}$
(Sec.~\ref{sec:framework}), so longitudinal two-momentum is \emph{not}
conserved on this background and plane waves are not eigenmodes of the
free evolution. The conserved label is the boost
(Schwarzschild-time/Rindler) frequency $\omega$ conjugate to the
dilatation generator, precisely the quantum number in which the elastic
near-horizon problem, the inverted-oscillator/dilatation picture
\cite{BGP2016,BGP2020}, is diagonal. We therefore expand per partial wave
with a generic discrete label $p\equiv(\omega,\text{channel})$,
\begin{equation}
  \hh^{\ell m}_{ab}(x)=\sum_p\Big[u^{(p)}_{ab}(x)\,b_{\ell m}(p)
  +u^{(p)*}_{ab}(x)\,\bdag_{\ell m}(p)\Big],
  \label{eq:modeexp}
\end{equation}
$[b,\bdag]$ canonical, with $u^{(p)}_{ab}$ the mode functions of the
quadratic operator normalized in the chosen vacuum. Two consequences are
used below: (i)~the momentum kernel of the vertex is the mode-function
overlap
\begin{equation}
  \mathcal K(p_1,p_2,p_3)\propto\int\dd^2x\;
  u^{(p_3)*}_{ab}\,\partial^au^{(p_1)}_{cd}\,\partial^bu^{(p_2)cd},
  \label{eq:kernel}
\end{equation}
which reduces to the plane-wave contraction
$p_{1a}p_{2b}\varepsilon_3^{ab}(\varepsilon_1\!\cdot\!\varepsilon_2)$ only
in the flat-space limit $\mu\to0$; and (ii)~the hierarchy of off-shellness that
orders the number-changing pieces is set by conservation of boost frequency,
$\omega_1+\omega_2=\omega_3$. The mode split
\eqref{eq:modeexp}, hence $b,\bdag$, the number operator, and the
inelasticity observable itself, is vacuum-dependent
(Boulware/Unruh/Hartle--Hawking); we adopt the GGV in/out near-horizon
basis, the basis in which the toolbox \cite{GGtoolbox} propagator is defined, and flag
this as a reported modeling input (Appendix~\ref{app:modeling}).

Inserting \eqref{eq:modeexp} into \eqref{eq:S3long}, each
$\hh\sim b+\bdag$ generates terms with $\Delta N\in\{\pm3,\pm1\}$. The
$\Delta N=\pm3$ pieces are the most off shell, their would-be resonance
$\omega_1+\omega_2+\omega_3=0$ is impossible for positive frequencies, so
$|\Delta\omega|\ge3\min_\ell\omega_\ell$, whereas the $\Delta N=\pm1$ pieces can
approach resonance ($\omega_1+\omega_2-\omega_3$ can be as small as the
$0.565/R_S$ gap of Sec.~\ref{sec:model}). We retain the least off-shell
$\Delta N=\pm1$ sector and drop the maximally off-shell $\Delta N=\pm3$ sector. This is a \emph{hierarchy truncation}, ordered by degree of off-shellness, rather than a rotating-wave approximation in the
standard sense: because the spectrum is resonance-free
(Sec.~\ref{sec:model}), \emph{every} retained process is itself off shell, so
there is no exactly-resonant subset that a genuine RWA would isolate. What
justifies the truncation is the parametric separation between the two off-shell
scales; the separation and its
$\ell$-dependence are quantified in Sec.~\ref{sec:selection}(c). The leading
number-changing-by-$\pm1$ piece is
\begin{equation}
  \boxed{\;
  H_3=\sum_{123}\Vg_{123}\big(\bdag_1\bdag_2b_3+\bdag_3b_2b_1\big)\;}
  \label{eq:H3}
\end{equation}
with $1\equiv(\ell_1m_1,p_1)$ etc.\ and
\begin{equation}
  \boxed{\;
  \Vg_{123}=\gamma\,W_{123}\,
  \Gaunt_{123}\,
  \mathcal K_{123}\,\mathcal N_{123}\;}
  \label{eq:Vgrav}
\end{equation}
where $\gamma=\kappa/\Rs$ (Finding~\ref{find:coupling}), $W_{123}\equiv
W(\lambda_1,\lambda_2,\lambda_3)$ is the
\emph{derived} on-shell polarization weight \eqref{eq:Wgen}
(Finding~\ref{find:cg}), $\Gaunt_{123}\equiv\Gaunt[\ell_1m_1;\ell_2m_2;\ell_3m_3]$
enforces \eqref{eq:msel}--\eqref{eq:paritysel},
$\mathcal K_{123}\equiv\mathcal K(p_1,p_2,p_3)$ is the kernel
\eqref{eq:kernel}, and $\mathcal N_{123}=(8\omega_1\omega_2\omega_3)^{-1/2}$
collects the mode normalizations. In the simulations below the boost
frequencies are carried per partial wave by the GGV dispersion
$\omega_\ell$ of Eq.~\eqref{eq:omegal}, one longitudinal oscillator per
$(\ell,m)$, and $\mathcal K=1$ in the leading soft limit, its variation
quantified in Sec.~\ref{sec:multil} as a systematic. On a single-$\ell$
register $W$ is the exact, independently confirmed constant $W(7,7,7)=-35.4375$
and is absorbed into the swept coupling $\gt$; across a multi-$\ell$ register the
derived ratios $W(\lambda_i)/W(7,7,7)$ set the physical relative weights, though
they carry the prescription systematic quantified in
Appendix~\ref{app:theoremcheck} (Sec.~\ref{sec:multil}).

\begin{remark}[$m$-conventions and Hermiticity: a trap for implementations]
\label{rem:mconv}
The Gaunt rule \eqref{eq:msel} applies to the harmonics of the
\emph{field} modes. Once the real field is expanded as
$\hh\sim\sum[b_{\ell m}Y^m_\ell+\bdag_{\ell m}Y^{m*}_\ell]$, each creation
operator enters with a conjugated harmonic; using
$Y^{m*}_\ell=(-1)^mY^{-m}_\ell$, the field-label rule translates into the
physical conservation law $m_1+m_2=m_3$ for the quanta in
$\bdag_1\bdag_2b_3$, with amplitude carrying the phase
$(-1)^{m_3}\Gaunt[\ell_1m_1;\ell_2m_2;\ell_3,-m_3]$ and a combinatorial
factor $2$ for distinct creation pairs. Transcribing \eqref{eq:msel} and
\eqref{eq:H3} literally, without the $(-1)^m$ relabeling, silently
violates $L_z$ conservation; Hermiticity pairs the kernel and
normalizations between the two terms of \eqref{eq:H3} and, with real Gaunt
coefficients and this phase convention, reduces to the reality of
$\Vg_{123}$. These properties are asserted at run time in every
simulation below.
\end{remark}

\begin{finding}[$H_3$ is number-changing with gravity-fixed coupling and
selection rules]
\label{find:H3}
The operator \eqref{eq:H3} changes graviton number by $\pm1$: one
longitudinal graviton splits into two and vice versa. Its coupling is
$\gamma$, and its selection rules are discrete
Wigner-$3j$ rules, exactly the two structural
properties whose absence would make any hand-chosen $V_{ijk}$ a
tautology.
\end{finding}

\section{The full simulation Hamiltonian and its exact structure}
\label{sec:model}

\subsection{Assembling $H$ and the Fock space}

The complete near-horizon Hamiltonian simulated below is
\begin{equation}
  \boxed{\;H=H_2^{\rm grav}+H_2^{\rm matter}
  +\Hint^{(h\phi\phi)}+H_3^{(hhh)}\;}
  \label{eq:Hfull}
\end{equation}
with $H_2=\sum_i\omega_{\ell_i}b_i^\dagger b_i$ per partial wave
[Eq.~\eqref{eq:omegal}], $\Hint$ the elastic vertex
\eqref{eq:Sintmatter} in second-quantized trilinear form
($b^\dagger_gaa+{\rm h.c.}$, same conjugation rule, reduced coupling
$\gtm$), and $H_3$ the derived vertex \eqref{eq:H3}--\eqref{eq:Vgrav} at
reduced coupling $\gt$: $\gt$ replaces the Planck-suppressed magnitude
$\gamma\,|W|\,\mathcal K$ and sweeps the overall scale of the
gravitationally \emph{fixed} relative structure, which is never dialed.
The black hole enters only as the fixed classical background; the quantum
states are Fock states of the partial-wave modes built on the chosen
vacuum, and no microstate is constructed. The explicit eleven-term
$\ell=2$ vertex table generated by \eqref{eq:Vgrav} is given in
Appendix~\ref{app:table}.

The physical role of $H_3$ in particle production is distinct from the
GGV exchange mechanism: matter feeds a graviton, the graviton
\emph{splits} (the new $H_3$ step), and each daughter reconverts to a
scalar pair,
\begin{equation}
  \phi\phi\xrightarrow{\;\Hint\;}h\xrightarrow{\;H_3\;}hh
  \xrightarrow{\;\Hint^2\;}\phi\phi\,\phi\phi,
  \label{eq:cascade}
\end{equation}
whereas the published $2\to2N$ tower \cite{GG2022} uses only $\Hint$,
gravitons exchanged between matter lines and never splitting. A single
$H_3$ insertion changes graviton number by one; matter multiplicity grows
by two per $h\to\phi\phi$ conversion.

\subsection{Exact selection rules}
\label{sec:selection}

Three structural statements control everything dynamical; all are
established analytically above and verified numerically at machine
precision.

\emph{(a) Magnetic rule and su(2) scalarity.} By
Remark~\ref{rem:mconv}, $[H,L_z]=0$ exactly (verified to $0$ in double
precision, Table~\ref{tab:validation}); forbidden $\Delta L_z=4$
amplitudes vanish identically; and the leading soft vertex is an exact
Wigner--Eckart su(2) scalar, the three independent $\ell=2$ witness
ratios agree to $O(10^{-18})$, with operator values
$\mathcal G_{\rm op}=+0.2207,\,-0.0901,\,+0.1802$ reproduced by the
simulation tables (Table~\ref{tab:validation}).

\emph{(b) The charge $Q$ and the opened channel.} The graviton--matter
model conserves
\begin{equation}
  Q=N_\phi+2N_h
  \label{eq:Q}
\end{equation}
under $H_2+\Hint$: we find $\|[H_2+\Hint,Q]\|=0$ exactly, while $H_3$
changes $Q$ by $\pm2$, $\|[H_3,Q]\|=3.52$ at $\gtthree=12$ on the register
of Sec.~\ref{sec:cascade}. Starting from two scalars ($Q=2$), the
four-scalar state ($Q=4$) is superselected away in the exchange-only
model and reachable only through the cascade \eqref{eq:cascade}, whose
minimal amplitude contains exactly one $H_3$ and three $\Hint$
insertions, fourth order, fixing the power laws tested in
Sec.~\ref{sec:cascade}.

\emph{(c) A resonance-free boost spectrum.} Because
$\Gaunt(\ell_1,\ell_2,\ell_3;0,0,0)\neq0$ exactly when parity and the
triangle inequality hold, the kinematically available cubic triads are
the parity- and triangle-allowed ones. Writing
$\omega_\ell=\sqrt{(\ell+\tfrac12)^2+\tfrac34}
=\ell+\tfrac12+\tfrac{3}{8(\ell+1/2)}+\dots$, the closest allowed triad
$\ell_3=\ell_1+\ell_2$ has mismatch
\begin{equation}
  \Delta\omega=\omega_{\ell_1}+\omega_{\ell_2}-\omega_{\ell_1+\ell_2}
  \;\longrightarrow\;\tfrac12^{\,+},
  \label{eq:gap}
\end{equation}
and smaller $\ell_3$ only increases it. A numerical audit over all
allowed triads with $\ell\le16$ gives $\min|\Delta\omega|=0.5653$ at
$(8,8)\to16$, monotonically approaching the bound $1/2$. \emph{No cubic
process in the near-horizon longitudinal sector is ever close to energy
shell.} This sharpens the status of the hierarchy truncation of
Sec.~\ref{sec:secondquant}: the near-resonance $\omega_1+\omega_2=\omega_3$
that would single out the retained $\bdag\bdag b$ form in a standard RWA is
itself never satisfied, the retained terms are off shell by at least $0.565$,
while the dropped $\Delta N=\pm3$ terms are off shell by
$\omega_1+\omega_2+\omega_3\ge3\sqrt7\simeq7.9$ on our registers, an order of
magnitude further. The truncation is thus justified by the parametric separation
between two off-shell scales; every
retained process acts virtually. Every consequence
reported below, perturbative onset, confirmed high-order power laws,
unit effective dimension, non-thermal late-time distributions, sub-Haar
entanglement, follows from this single spectral fact combined with the
bounded vertex magnitudes $|V_{123}|\le0.0363\,\gt$
(Appendix~\ref{app:table}): at the representative $\gt=12$ the largest
vertex element $0.435$ remains well below the smallest denominator
$\Delta\omega=\sqrt7\simeq2.65$ of the $\ell=2$ register, and the
measured inelasticity $\eta\simeq0.10$--$0.13$ is consistent with the
perturbative estimate $\sum_c(V_c/\Delta\omega)^2$ over the $O(4)$ open
channels.

\section{Simulation framework and validation}
\label{sec:methods}

\subsection{Encodings and registers}

Bosonic modes with per-mode cutoff $d$ are mapped to $n_b=\lceil\log_2
d\rceil$ qubits per mode in standard-binary, Gray \cite{Sawaya2020,
DiMatteo2021}, or unary encodings; all production runs use Gray, $d=4$
($n_b=2$). The $\ell=2$ multiplet (5 modes, 10 qubits) yields a
1103-term \texttt{SparsePauliOp}; the 3-mode subset (6 qubits), 283 terms.
Encodings are validated bit-for-bit against the dense ladder algebra for
$d=2,3,4,8$ (Table~\ref{tab:validation}).

\subsection{Two truncation schemes and the symmetry-preserving cut}
\label{sec:truncation}

The per-mode cutoff $d$ is what a qubit register implements, but it breaks
the model's su(2) at the occupation boundary: we measure
$\|[H,L^2]\|=22.16$ at $d=4$ on the $\ell=2$ multiplet. We therefore
introduce, as an exact-diagonalization companion, the \emph{total-occupation}
truncation $\sum_i n_i\le \Nmax$: since the ladder generators $L_\pm$
conserve total $N$, this subspace is exactly su(2)-invariant, and we verify
$\|[H,L^2]\|=8.9\times10^{-16}$ (Table~\ref{tab:validation}). All
symmetry-resolved spectral statements (Sec.~\ref{sec:spectral}) and the
multi-$\ell$ convergence studies use this cut. One practical remark spares
future implementations a silent error: within an $L_z$-restricted basis the
Casimir must be assembled as the \emph{composite} operator $L_+L_-$ summed
through intermediate configurations outside the sector; the product of
sector-restricted ladder matrices vanishes identically.

\subsection{Circuits, evolution, and measurement}
\label{sec:circuits}

Time evolution uses second-order Suzuki--Trotter product formulas
\cite{Childs2021}. Every \texttt{PauliEvolutionGate} is explicitly
decomposed before simulation, applying \texttt{Statevector} to the
undecomposed gate silently substitutes the exact exponential and masks all
Trotter error; we verify decomposition-independence at $6.7\times10^{-13}$
and a genuine global $r^{-4}$ error scaling (fitted slope $-4.03$;
infidelity $1.65\times10^{-2}\to2.08\times10^{-8}$ for $r=1\to32$ on the
elastic benchmark, Appendix~\ref{app:elas-bench}). Long evolutions use a
sequential-segment protocol: the transpiled segment $U_T(\Delta t)$ is
applied repeatedly on the Aer statevector backend, feeding each saved state
back as the next initial state, so memory is bounded by one segment and the
Trotter error accumulates linearly as it would on hardware. Multiplicity
measurement is computational-basis readout decoded to occupations;
$P(N,t)$ estimates carry binomial errors. The squared commutator
$C(t)=\langle|[W(t),V]|^2\rangle$ \cite{MSS2016} is measured by an ancilla
Hadamard test on the controlled conjugated sequence, with Pauli probes on
physical qubits (Appendix~\ref{app:protocols}).

\subsection{The quantum circuits}
\label{sec:circuitfigs}

\begin{figure}[t]
\includegraphics[width=\columnwidth]{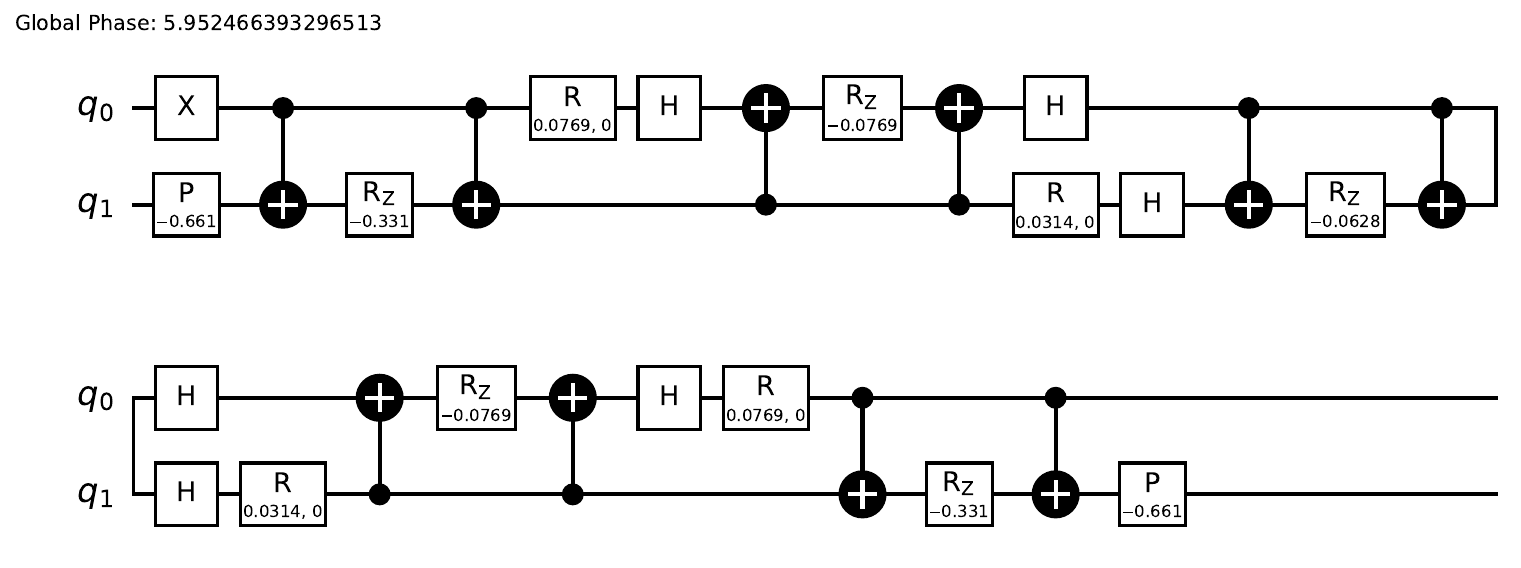}
\caption{\label{fig:Q1}The complete circuit for the smallest instance of
the derived model: Fock preparation plus one decomposed second-order
Suzuki--Trotter step for the $(2,0)$ mode at $d=4$ (Gray encoding, two
qubits, seven Pauli strings; depth 25, 10 cx). Production registers scale
this construction to 6 and 10 qubits (Table~\ref{tab:resources}).}
\end{figure}

\begin{figure}[t]
\includegraphics[width=\columnwidth]{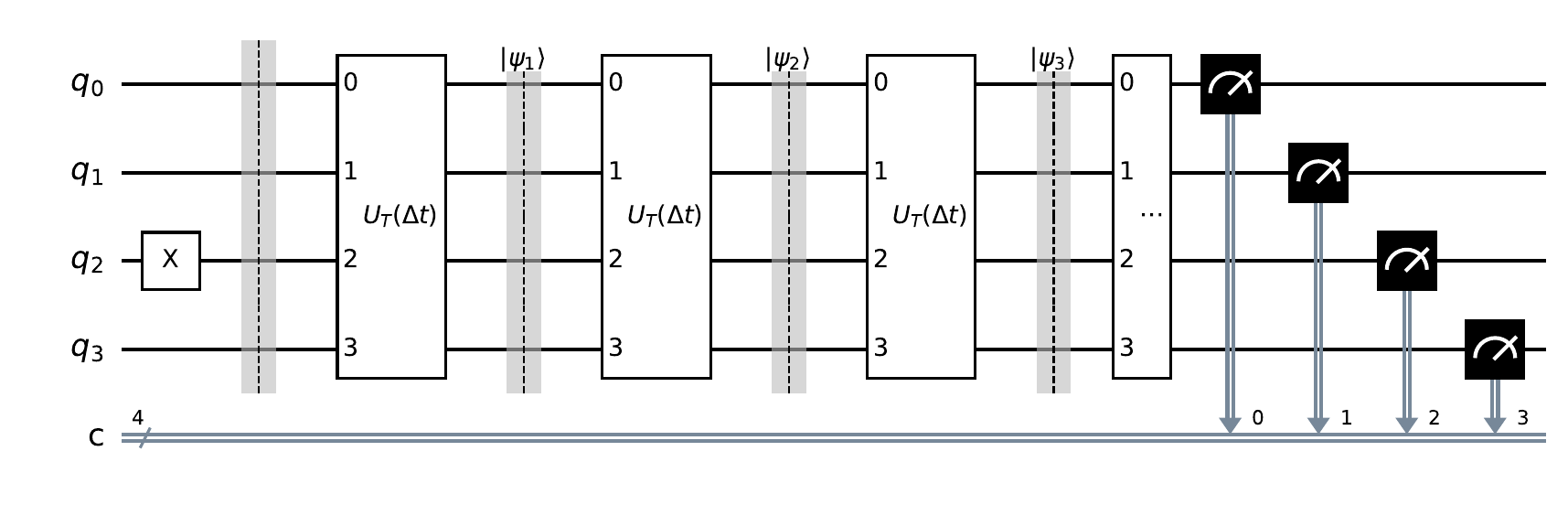}
\caption{\label{fig:Q2}The sequential-segment snapshot protocol:
$X$-layer Fock preparation, repeated transpiled Trotter segments
$U_T(\Delta t)$ with the state recorded after each segment
($|\psi_k\rangle$ at $t=k\Delta t$), and computational-basis
multiplicity readout. Memory is bounded by one segment; the Trotter error
accumulates linearly, as on hardware.}
\end{figure}

\begin{figure}[t]
\includegraphics[width=\columnwidth]{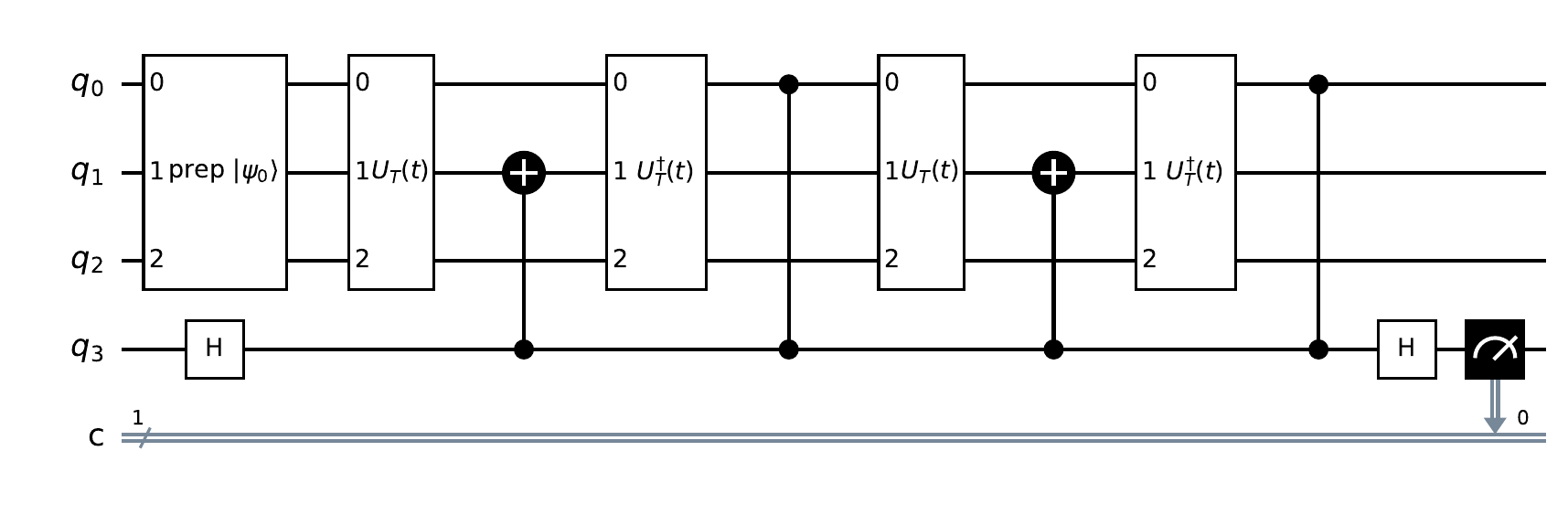}
\caption{\label{fig:Q3}The Hadamard-test interferometer measuring the
squared commutator with Pauli probes ($V=X$, $W=Z$ shown): controlled
time evolution is avoided by conjugation,
${\rm c}\,(U^\dagger VU)=U\,{\rm c}V\,U^\dagger$, so only the
single-qubit probes are controlled; the final ancilla
$\langle Z\rangle$ gives ${\rm Re}\,F$ and $C=2-2\,{\rm Re}\,F$.}
\end{figure}

Figures~\ref{fig:Q1}--\ref{fig:Q3} display the three circuit primitives
of the pipeline at readable scale; all are drawn from the production code. Figure~\ref{fig:Q1} is the complete, real
circuit for the smallest instance of the derived model: the single mode
$(2,0)$ at $d=4$ in Gray encoding (two qubits), whose Hamiltonian
$\omega\,b^\dagger b+V_{(000)}(\bdag\bdag b+{\rm h.c.})$ maps to seven
Pauli strings, $H_2$ generating the diagonal $ZI/ZZ$ rotations and the
derived self-term $(2,0)^3$ (allowed by the selection rules, since
$m$: $0+0=0$ and parity $2+2+2$ even) generating the off-diagonal
$IX,ZX,XI,XZ$ rotations. One second-order Suzuki--Trotter step decomposes
to depth 25 with 10 cx. The production registers are the same construction
scaled up (Table~\ref{tab:resources}). Figure~\ref{fig:Q2} shows the
sequential-segment snapshot protocol of Sec.~\ref{sec:circuits}: an
$X$-layer Fock preparation followed by repeated transpiled segments
$U_T(\Delta t)$ with the state saved (and, on hardware, measured) after
each segment. Figure~\ref{fig:Q3} is the ancilla Hadamard-test
interferometer for the squared commutator, faithful to the implementation:
controlled evolution is avoided by conjugation, so only the Pauli probes
$V,W$ are controlled, the sequence $[U_T,{\rm c}V,U_T^\dagger,{\rm c}W]$
is applied twice, and $\langle Z_{\rm anc}\rangle={\rm Re}\,F$ with
$C=2-2\,{\rm Re}\,F$.

\subsection{Validation stack}

Table~\ref{tab:validation} consolidates the run-time checks that gate every
production result. The elastic benchmark (Appendix~\ref{app:elas-bench}) additionally
calibrates the full circuit machinery on exactly solvable near-horizon
physics: the inverted-oscillator squared commutator reproduces
$\cosh^2\Omega t$ to a relative $1.7\times10^{-10}$, the exact 't~Hooft
shift map $S=\exp(ic\,p_{\rm in}p_{\rm out})$ \cite{DrayTHooft1985,
tHooft1996} is implemented as a cross-register diagonal circuit exact to
$5.0\times10^{-15}$, and the Lyapunov-fit pipeline recovers the analytic
growth rate $2\Omega$ to $0.2\%$ \emph{on the benchmark where exponential
growth exists}, which is what licenses the bounded-OTOC null result of
Sec.~\ref{sec:otoc}.

\begin{table}[t]
\caption{\label{tab:validation}Validation stack ($\ell=2$ multiplet,
$d=4$, $\gt=12$ where applicable). E1/E2/E3 denote the dense-Kronecker,
occupation-basis, and qubit-Pauli engines.}
\begin{ruledtabular}
\begin{tabular}{ll}
Check & Value \\
\hline
Vertex terms, $\ell=2$ multiplet & 11 \\
$\max|{\rm E1}-{\rm E2}|$ & $7.1\times10^{-15}$ \\
$\max|{\rm E1}-{\rm E3}|$ (std.~bin.) & $1.8\times10^{-14}$ \\
E1 vs E3 spectrum (Gray) & $1.6\times10^{-13}$ \\
Pauli terms (5 modes, 10 qubits) & 1103 \\
$\|[H,L_z]\|$ & $0$ \\
$\|[H,P_{m\to-m}]\|$ & $7.1\times10^{-15}$ \\
$\|[H,L^2]\|$, per-mode $d{=}4$ & $22.16$ \\
$\|[H,L^2]\|$, total-$N$ cutoff & $8.9\times10^{-16}$ \\
$\mathcal G_{\rm op}$ witnesses & $0.2207,\,-0.0901,\,0.1802$ \\
Forbidden $\Delta L_z=4$ amplitude & $0$ \\
\end{tabular}
\end{ruledtabular}
\end{table}

\section{Results}
\label{sec:results}

\subsection{Inelasticity onset and multiplicity dynamics}
\label{sec:onset}

\begin{figure}[t]
\includegraphics[width=\columnwidth]{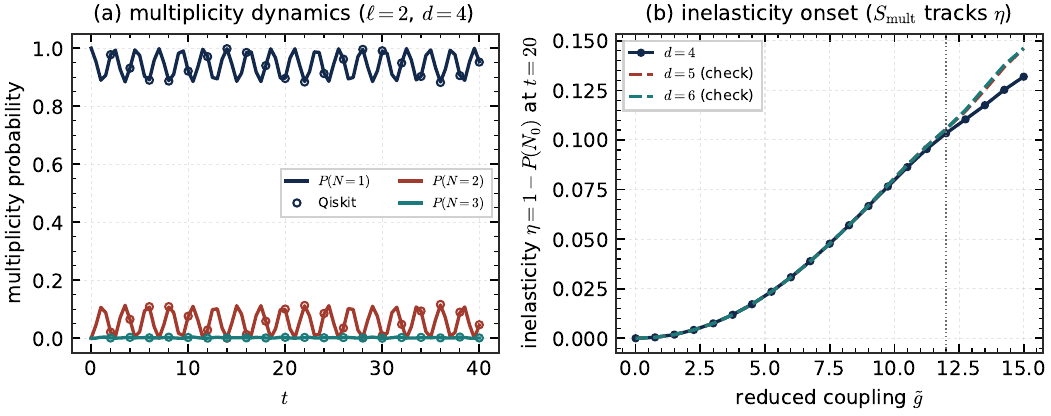}
\caption{\label{fig:B1}(a)~Multiplicity dynamics $P(N,t)$ at $\gt=12$
($\ell=2$ multiplet, $d=4$): exact block evolution (lines) and the 10-qubit Gray
Trotter engine (open circles; $\max$ deviation $1.6\times10^{-3}$).
(b)~Inelasticity onset $\eta(\gt)$ at $t=20$ for $d=4,5,6$, with the multiplicity
entropy $S_{\rm mult}$ tracking $\eta$; the dotted line marks the representative
coupling $\gt=12$.}
\end{figure}

Figure~\ref{fig:B1} shows the graviton multiplicity $P(N,t)$ on the
$\ell=2$ multiplet ($d=4$, $L_z=0$ block) from a single $(2,0)$ graviton at
the representative coupling $\gt=12$, computed three ways: exact block
evolution, the 10-qubit Gray-encoded Trotter engine (segment protocol,
$r=16$ per unit $\Delta t=2$), and Aer sampling with $16384$ shots at
$t=20$. The Trotter engine tracks the exact curves to
$\max|\Delta P|=1.6\times10^{-3}$ across all twenty snapshots; the sampled
estimates $P_1=0.8878(25)$, $P_2=0.1086(24)$, $P_3=0.0034(5)$,
$P_4=2.4(1.2)\times10^{-4}$ agree with the exact values
$(0.8966,\,0.0996,\,0.0034,\,3.1\times10^{-4})$ within the combined shot
error and the $O(10^{-2})$ Trotter error of the shallower ($r=60$) sampling
circuit. The maxima over the window are $\eta_{\max}=0.1167$ and
$P_2^{\max}=0.1150$.

Figure~\ref{fig:B1}(b) gives the onset $\eta(\gt)$ and multiplicity entropy
$S_{\rm mult}(\gt)$ at $t=20$ with cutoff overlays $d=4,5,6$: the maximal
truncation shift is $|\eta_{d=4}-\eta_{d=6}|\le 0.0144$ over the full sweep
($\eta=0.132$ vs $0.146$ at the sweep edge $\gt=15$), so the working window
$\gt\lesssim12$--$15$ is Fock-converged at the few-percent level.

\subsection{Late-time distribution: four references and an effective
dimension of one}
\label{sec:references}

\begin{figure}[t]
\includegraphics[width=\columnwidth]{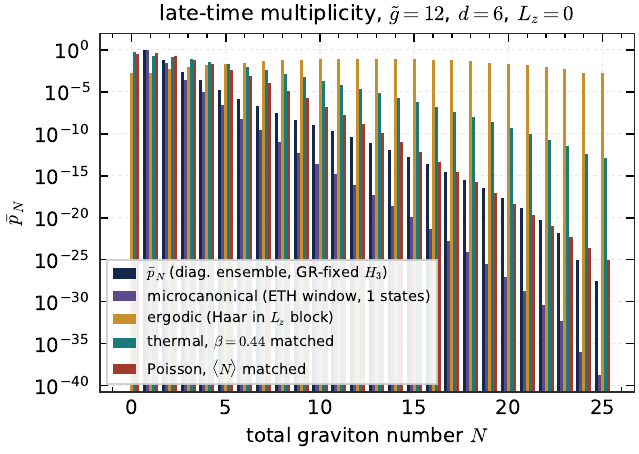}
\caption{\label{fig:B3}Late-time multiplicity $\bar p_N$ (diagonal ensemble,
$\gt=12$, $d=6$, $L_z=0$) against the microcanonical window, Haar-block,
$\beta$-matched thermal, and Poisson references. The distribution is far
from every ergodic reference; the closeness to the single-state
microcanonical window is the statement $d_{\rm eff}=1.06$.}
\end{figure}

The infinite-time (diagonal-ensemble \cite{Rigol2008}) distribution
$\bar p_N$ at $\gt=12$ on the largest converged register ($d=6$, $L_z=0$) is
compared in Fig.~\ref{fig:B3} against four references computed in the same
block: the microcanonical (ETH-window \cite{Deutsch1991,Srednicki1994,
DAlessio2016}) eigenstate average over the energy shell actually explored;
the Haar-typical (Page \cite{Page1993}) block average; the
$\beta$-matched Gibbs distribution of $H_2$ ($\beta=0.440$); and the
$\langle N\rangle$-matched Poisson distribution. The measured distribution,
$\bar p_N=\{0.9415,\,0.0563,\,0.0020,\,2.4\times10^{-4},\dots\}$ for
$N=1,2,3,4$ with $\langle N\rangle=1.061$, is quantitatively far from all
ergodic references:
\begin{align}
  {\rm TV}({\rm thermal})&=0.777,\quad
  {\rm TV}({\rm Poisson})=0.574,\nonumber\\
  {\rm TV}({\rm Haar})&=0.990,
  \nonumber
\end{align}
(KL divergences $1.58$, $0.81$, $6.01$). The closeness to the microcanonical window (${\rm TV}=0.028$) reflects the single-eigenstate content of the energy shell rather than eigenstate thermalization: $E_0\pm\sigma_E$ ($\sigma_E=0.470$) contains a single eigenstate at
both $1\sigma$ and $2\sigma$ width, and the initial state's effective
dimension over energy eigenstates is
\begin{equation}
  d_{\rm eff}\;=\;\Bigl(\sum_n|c_n|^4\Bigr)^{-1}\;=\;1.062 .
  \label{eq:deff}
\end{equation}
The quench does not delocalize: the single graviton becomes a single
\emph{dressed} eigenstate whose $N$-content is the measured $\bar p_N$.
This is the quantitative meaning of perturbative rigidity, and it is the
behavior the resonance-free spectrum [Eq.~\eqref{eq:gap}] mandates at
couplings within the Fock-converged window.

\subsection{Structure ablation: what is gravitationally specific}
\label{sec:ablation}

\begin{figure}[h!]
	\includegraphics[width=\columnwidth]{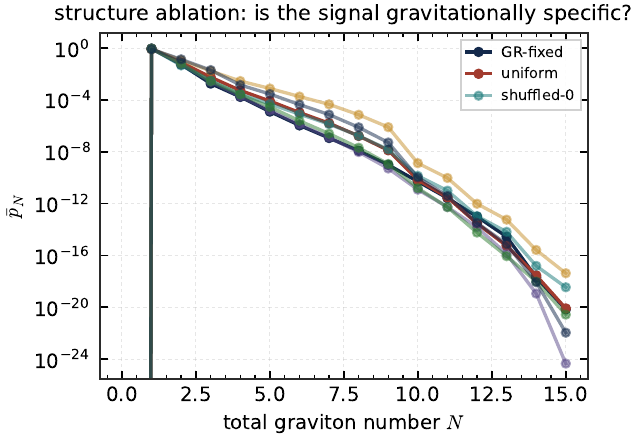}
	\caption{\label{fig:B4}Structure ablation of $\bar p_N$: gravitationally
		fixed table versus uniform-magnitude and five magnitude-shuffled variants
		(selection rules preserved). $\eta$ varies ninefold across variants while
		all remain non-ergodic.}
\end{figure}

Is the measured signal a property of the gravitationally fixed structure,
or of any cubic model with these frequencies? We rerun the identical quench
with (i)~all vertex magnitudes replaced by their uniform mean and (ii)~five
magnitude-reshuffled tables (selection rules intact, seeds $0$--$4$). The
inelasticity is strongly structure-sensitive: $\eta=0.1034$ (GR-fixed)
versus $0.1349$ (uniform) and $\{0.0142$--$0.1194\}$ across shuffles, a
ninefold spread under reshuffling of the same magnitude multiset
(Fig.~\ref{fig:B4}). The \emph{shape} of $\bar p_N$ is more robust
[${\rm TV}({\rm GR,shuffled})=0.009$--$0.104$, mean $0.043$], and every
variant remains equally non-ergodic (each within ${\rm TV}\le0.075$ of its
own single-state window). The decomposition is clean: non-thermality is generic to the resonance-free spectrum, while the magnitude and channel weighting of the inelasticity carry the gravitational fingerprint.

\subsection{Multi-$\ell$ register under the symmetry-exact truncation}
\label{sec:multil}

\begin{figure}[t]
\includegraphics[width=\columnwidth]{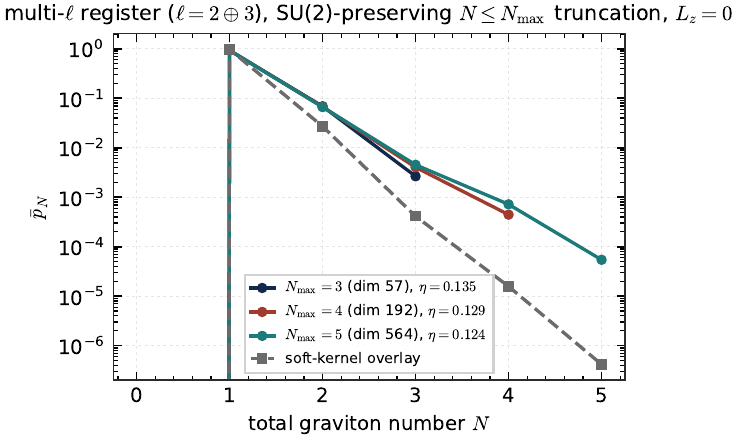}
\caption{\label{fig:B5}Late-time $\bar p_N$ on the $\ell=2\oplus3$ register
under the su(2)-exact total-$N$ truncation, $\Nmax=3,4,5$, with the
\emph{derived}-weight comparison overlay: flat leading-soft $K=1$ versus the
derived on-shell weight $W(\lambda_1,\lambda_2,\lambda_3)$ at matched
Frobenius norm (${\rm TV}=0.093$; $\eta_{\rm late}$ shift $-8.6\%$).}
\end{figure}

On the twelve-mode $\ell=2\oplus3$ register with the total-$N$ cut
(Sec.~\ref{sec:truncation}; $[H,L^2]=0$ at machine precision throughout,
block dimensions $57/192/564$ for $\Nmax=3/4/5$), the inelasticity at $t=20$ is
$\eta=0.1354,\,0.1285,\,0.1235$, monotonically converging at the
$4\%$ level per step (Fig.~\ref{fig:B5}). Here the derived polarization
weight $W$ has genuine content: on this multi-$\ell$ register the ratios
$W(\lambda_2)/W(\lambda_3)$ are fixed numbers (Sec.~\ref{sec:theorem},
Appendix~\ref{app:table}). Replacing the flat leading-soft
weight $K=1$ by the derived $W(\lambda_1,\lambda_2,\lambda_3)$, norm-matched
to the same Frobenius scale so only the \emph{relative} multipole structure
differs, shifts the late-time longitudinal-occupation distribution by
${\rm TV}=0.093$ on $\bar p_N$ and the inelasticity from
$\eta_{\rm late}=0.774$ ($K=1$) to $0.708$ (derived), a $-8.6\%$ change.
This is a \emph{derived} systematic, the physical multipole weighting
of the vertex. The single-$\ell$
observables are unaffected, since there $W$ is an exact overall constant
absorbed into $\gt$.

\subsection{Entanglement profile: 2.3 nats below Haar}
\label{sec:page}

\begin{figure}[t]
\includegraphics[width=\columnwidth]{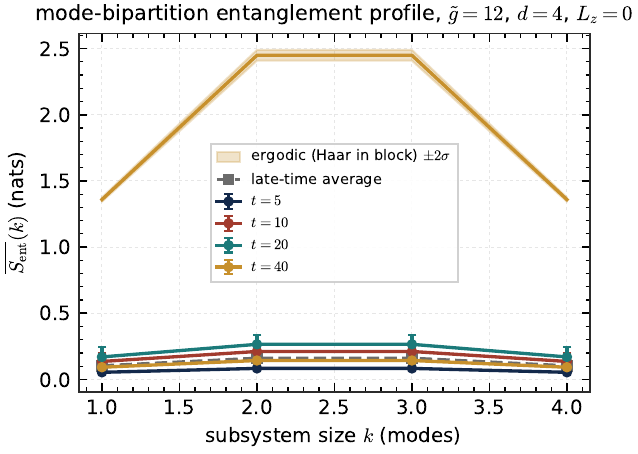}
\caption{\label{fig:C1}Mode-bipartition entanglement profile at several
times versus the Haar (Page) reference band in the same block ($\gt=12$,
$d=4$, $L_z=0$). The late-time average sits $2.29$ nats below the Haar
half-cut value.}
\end{figure}

Figure~\ref{fig:C1} shows the mode-bipartition entanglement profile
$\overline{S_{\rm ent}}(k)$ of the evolving state against the Haar-typical
(Page \cite{Page1993}) reference computed in the same $L_z$ block. The late-time-averaged
profile, $\{0.103,\,0.160,\,0.160,\,0.103\}$ nats for $k=1..4$, sits far
below the Haar band $\{1.359,\,2.449,\,2.449,\,1.359\}$: the maximal
deficit is $2.29$ nats at half-cut. Growth of $\eta$, $S_{\rm mult}$, and
$S_{\rm ent}$ is synchronized and oscillatory rather than saturating
(Appendix~\ref{app:supp}), and the multi-$\ell$ profile under the
symmetry-exact cut is likewise sub-Haar and $\Nmax$-convergent
(Appendix~\ref{app:supp}). The produced state is weakly and structuredly
entangled.

\subsection{The channel that only $\Hthree$ opens: superselection and
confirmed power laws}
\label{sec:cascade}

\begin{figure*}[t]
\includegraphics[width=0.85\textwidth]{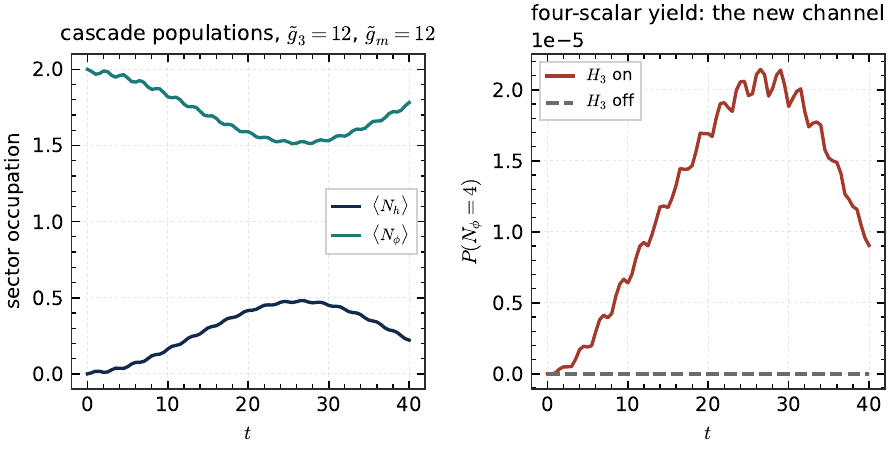}
\caption{\label{fig:F1}The $\Hthree$-opened cascade
[Eq.~\eqref{eq:cascade}] on the mixed graviton--scalar register: sector
populations (left) and the four-scalar yield with the vertex on versus off
(right). The off curve is an exact zero by the superselection of
Eq.~\eqref{eq:Q}.}
\end{figure*}

On the mixed register of five $\ell=2$ gravitons and three $\ell=2$ scalars
($d=3$, $L_z=0$ block, dimension 891), initialized with the scalar pair
$\phi_{(2,1)}\phi_{(2,-1)}$, the four-scalar probability with $\Hthree$ off
is zero to numerical precision ($\le 2\times10^{-31}$), as the charge
[Eq.~\eqref{eq:Q}] requires; with the derived vertex on it rises to
$P_4^{\max}=2.14\times10^{-5}$ at $\gtthree=\gtm=12$
(Fig.~\ref{fig:F1}). The on/off contrast is fixed by the superselection rule; the quantitative content is the absolute yield, its coupling dependence, and its systematics. The yield at $\gtthree=12$ carries a
$21\%$ per-mode-truncation shift ($1.69\times10^{-5}$ at $d=3$ vs
$1.40\times10^{-5}$ at $d=4$): absolute yields are order-accurate, while
the \emph{exponents} below are truncation-robust.

\begin{figure*}[t]
\includegraphics[width=0.98\textwidth]{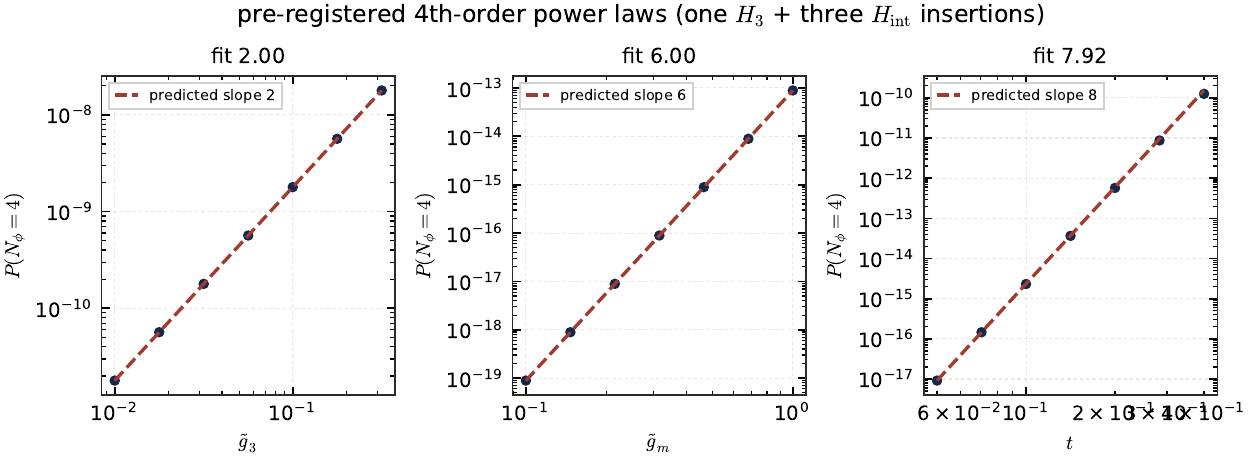}
\caption{\label{fig:F3}Pre-registered fourth-order power laws of the opened channel, with fitted exponents $2.000$, $6.000$, $7.92$ against predictions $2$, $6$, $8$ (dashed).}
\end{figure*}

The minimal fourth-order amplitude (one $\Hthree$, three $\Hint$) fixes
three leading-order scaling laws; these are perturbative consistency checks
of the simulation, linear (and low-order) response with known exponents,
stated in advance and confirmed by log--log fits (Fig.~\ref{fig:F3}):
\begin{equation}
  P_4\propto \gtthree^{\,2}\!:2.000,\quad
  P_4\propto \gtm^{\,6}\!:6.000,\quad
  P_4\propto t^{8}\!:7.92,
  \label{eq:powerlaws}
\end{equation}
the residual in the last reflecting the known $O(t)$ subleading correction
of the short-time expansion. The $\gtthree^{\,2}$ law is exact linear
response in $\Hthree$ and holds at strong $\gtm$, as measured.

\subsection{Spectral statistics, two-sided}
\label{sec:spectral}

The level statistics of the model require the same discipline as its
dynamics, because the two truncations of Sec.~\ref{sec:truncation} give
qualitatively different answers, and only one of them is attributable to
the physics.

\begin{figure*}[t]
\includegraphics[width=0.45\textwidth]{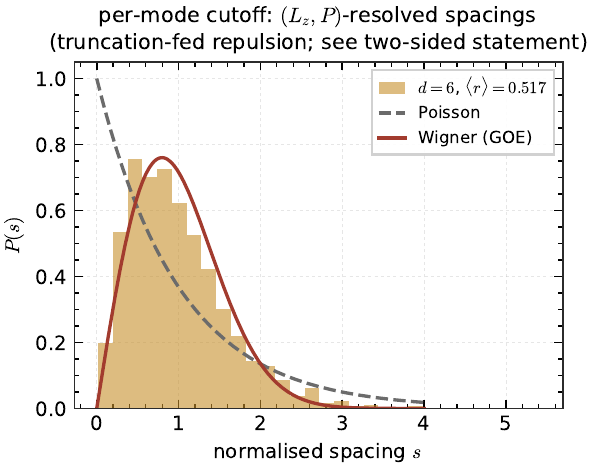}
\includegraphics[width=0.48\textwidth]{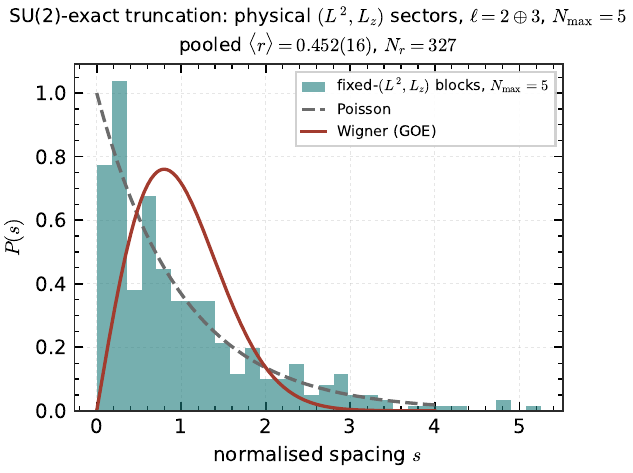}
\caption{\label{fig:D1}Two-sided spectral statistics. Left:~Under the per-mode
occupation cutoff the $(L_z,P)$-resolved level spacings show GOE-grade repulsion
that is truncation-fed. Right:~Under the su(2)-exact total-occupation truncation the
physical $(L^2,L_z)$ sectors show intermediate statistics, $\rmean=0.452(16)$,
between Poisson and Wigner.}
\end{figure*}

\emph{Per-mode cutoff (the encoded Hamiltonian).} Resolving the exact
symmetries that survive the cutoff ($L_z$ and $m\to-m$ parity) and pooling
the $(L_z,P)$ blocks, the mean gap ratio \cite{OganesyanHuse2007,Atas2013}
is refinement-stable GOE-grade:
$\rmean=0.4814\,(359\ {\rm spacings}),\ 0.5131\,(938),\ 0.5168\,(2000)$ for
$d=4,5,6$, against ${\rm GOE}=0.5307$ and ${\rm Poisson}=0.3863$
(Fig.~\ref{fig:D1}). This repulsion is fed by the cutoff-induced su(2)
breaking ($\|[H,L^2]\|=22$) and cannot be attributed to the untruncated
dynamics.

\emph{Symmetry-exact truncation (the physical sectors).} Under the
total-$N$ cut the spectrum resolves into genuine fixed-$(L^2,L_z)$ sectors
(seven to sixteen sectors across our registers; $\|[H,L^2]\|\le
2\times10^{-13}$ throughout). On the single $\ell=2$ multiplet every
physical sector remains too small for spacing statistics even at $\Nmax=7$
(dimensions $18$--$72$ split across up to thirteen sectors). On the $\ell=2\oplus3$ register
the sectors are large enough, and the weighted mean over all sectors with
$\ge10$ spacings is
\begin{align}
  \rmean &= 0.351\ (\Nmax{=}4,\ 81\ {\rm spacings})\nonumber\\
   &\to\; 0.452(16)\ (\Nmax{=}5,\ 347\ {\rm spacings}),
  \label{eq:rphys}
\end{align}
with per-sector values $0.36$--$0.50$ across $L=0$--$10$
(Fig.~\ref{fig:D1}). The physical-sector statistics are \emph{indicative} of
intermediate statistics, $4.1\sigma$ above Poisson, $5.0\sigma$ below GOE, and
rising with $\Nmax$ at the accessible depths. The individual sectors ($\ge10$ spacings each, largest a few hundred)
remain far too small to \emph{establish} a universality class, so we report an indicative trend. Whether the physical model becomes chaotic at larger $\Nmax$ is
an open, now sharply posed, question.

\subsection{Out-of-time-order correlators: bounded, with a validated
instrument}
\label{sec:otoc}

\begin{figure*}[t]
	\includegraphics[width=0.9\textwidth]{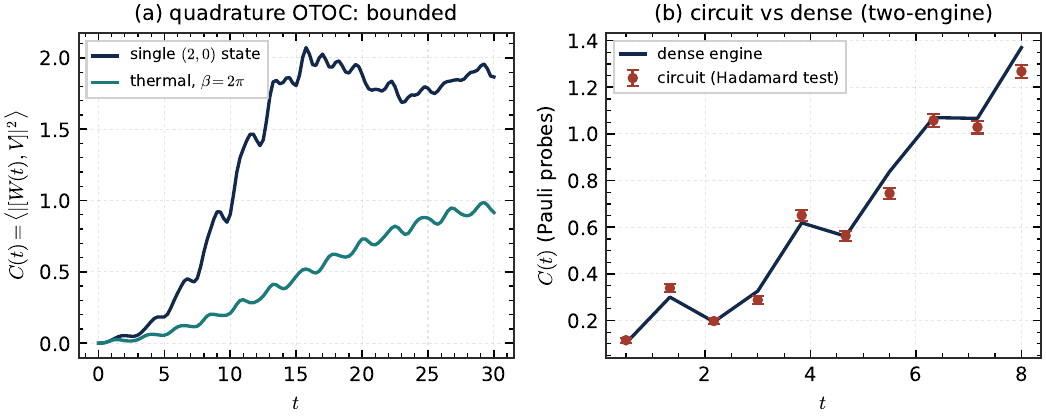}
	\caption{\label{fig:E1}(a)~Quadrature squared commutator on the $\ell=2$
		multiplet for a single-graviton state and the $\beta=2\pi$ thermal state: bounded
		and non-scrambling at accessible sizes.
		(b)~The circuit protocol (ancilla Hadamard test, Pauli probes, $16384$ shots)
		against the dense engine, a two-engine validation.}
\end{figure*}

The squared commutator between quadratures of distinct multiplet modes is
bounded over the full window for both a single-graviton state
($C_{\max}=2.07$) and the horizon-temperature thermal state $\beta=2\pi$
($C_{\max}=0.99$), with no clean exponential window at accessible sizes
(Fig.~\ref{fig:E1}); accordingly \emph{no Lyapunov exponent is extracted
and no saturation of the chaos bound \cite{MSS2016} is claimed}. The null
result is instrumented: the identical fitting pipeline applied to the inverted-oscillator benchmark, where $C=\cosh^2\Omega t$ exactly,
recovers the analytic squared-commutator rate $2\Omega$ to $0.2\%$
\cite{Polchinski2015}, and the hardware-format ancilla Hadamard test
reproduces the dense $C(t)$ at ten time points within $0.10$ absolute on
$C\in[0,1.37]$, dominated by the Trotter error of the chosen depth, with
mean $2\sigma$ shot error $0.021$ at $16384$ shots
(Appendix~\ref{app:supp}). A bounded squared commutator at effective dimension $d_{\rm eff}\simeq1$ is what a rigid sector and a merely-small Hilbert space both produce, so this observable becomes decisive only at larger register size, which is exactly what the symmetry-exact truncation is built to reach. The tensor-network climb (Sec.~\ref{sec:mps}) finds the non-scrambling behavior extends, area-law, to sixty modes.

\subsection{Tensor-network climb: the multipole ladder to sixty modes}
\label{sec:mps}

The registers above span one or two multiplets because the cost of exact
diagonalization and statevector simulation grows exponentially in the mode
number. The derived Hamiltonian is, however, a mode Hamiltonian whose only
exactly conserved abelian charge is the azimuthal weight $L_z=\sum_i m_i n_i$.
Grading a matrix-product-operator representation of $H_2^{\rm grav}+\Hthree$ by
this $U(1)$ charge, so that the cubic vertex $\bdag_i\bdag_j b_k$ is admitted
only when $m_i+m_j=m_k$, lets the single-$(2,0)$-graviton quench be evolved by
tensor-network methods far beyond the reach of the statevector engine. We climb
the full ladder $\ell_{\max}=2,\dots,7$, that is $M=5,12,21,32,45,60$ modes,
the last row being the $120$-qubit register of Table~\ref{tab:scaling}. The
engine, its validation against the exact-diagonalization Hamiltonian to machine
precision, and the convergence controls are given in
Appendix~\ref{app:scaling:climb}.

\begin{figure}[t]
\includegraphics[width=\columnwidth]{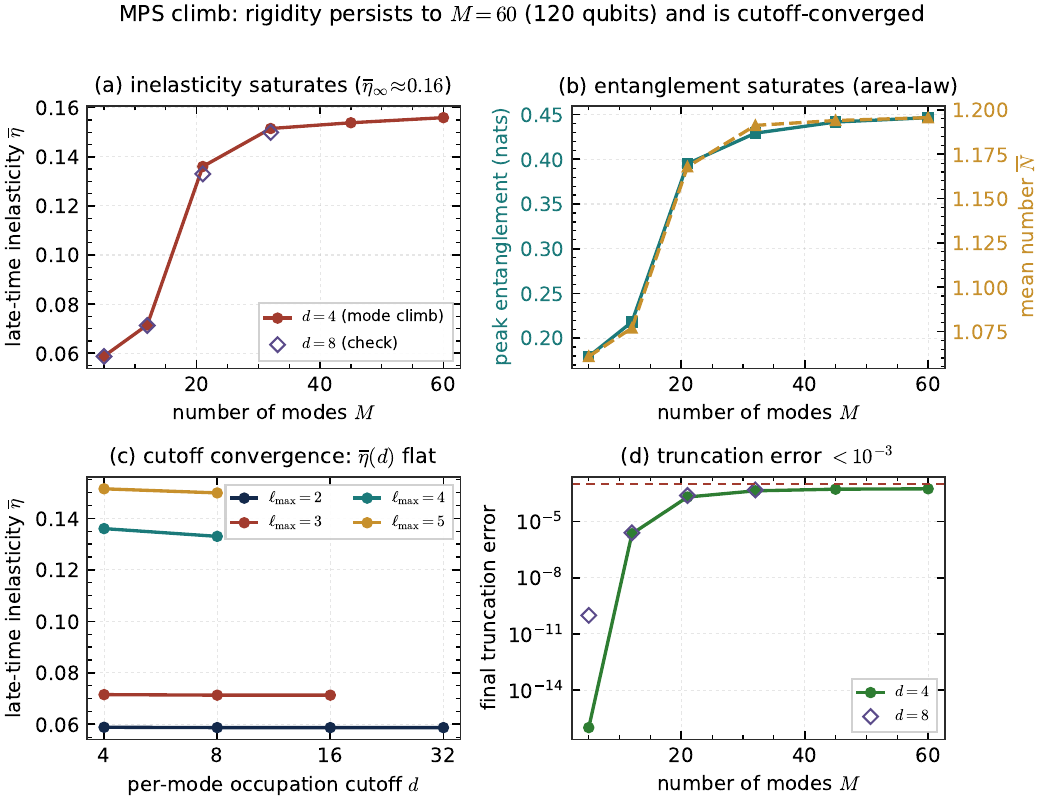}
\caption{\label{fig:H2}Tensor-network climb of the multipole ladder to $M=60$
modes ($120$ qubits). (a)~The late-time inelasticity saturates at
$\bar\eta_\infty\simeq0.16$. (b)~The peak mode-bipartition entanglement
$\overline{S}_{\max}$ and the mean occupation $\bar N$ both saturate; the peak
of $\simeq0.45$ nats lies far below the Page value for sixty modes (area-law).
(c)~Cutoff convergence: the late-time inelasticity $\bar\eta(d)$ is flat across
$d=4$--$32$ at each depth (the occupation saturates; $d=8$, $16$ and $32$
coincide at $\ell_{\max}=2,3$). (d)~The final truncation error stays below the
Trotter-level $10^{-3}$ at every register. Open diamonds in (a) and (d) are the
$d=8$ cross-check, which reaches $96$ qubits at $\ell_{\max}=5$. Single-$(2,0)$
quench, $\gt=12$, $L_z=0$, bond dimension $\chi=24$.}
\end{figure}

Three quantities saturate as the multipole tower deepens (Fig.~\ref{fig:H2}).
The inelasticity rises from the two-multiplet value and levels at
$\bar\eta_\infty\simeq0.156$ by $M\gtrsim32$; the mean longitudinal occupation
levels at $\bar N\simeq1.20$; and the peak mode-bipartition entanglement levels
at $\overline{S}_{\max}\simeq0.45$ nats. The last number is the decisive one:
a Haar-typical state on sixty modes would carry tens of nats across a central
cut, so a peak of half a nat is an area-law
dressing. The entanglement is moreover localized: its
late-time profile peaks near the active low-$|m|$ modes and decays across the
chain, with the profiles for $\ell_{\max}=4$ through $7$ collapsing onto a
common envelope (Appendix~\ref{app:scaling:climb}, Fig.~\ref{fig:H1}). The
perturbative rigidity established analytically and at $\ell=2$ therefore
persists, quantitatively and without qualitative change, to the full
$120$-qubit register.

The climb settles the inelasticity on a measure common to both engines.
For the cross-engine comparison and the multipole convergence we report the
late-time average of $\eta(t)=1-P(N_0,t)$ over the accessible window, which washes out the quasi-periodic oscillation; computed
this way from the \emph{same} Qiskit multiplicity trajectory, the
single-multiplet inelasticity is $\bar\eta\simeq0.058$ at $\ell=2$, in agreement
with the matrix-product-state value, whereas the frequently quoted single
snapshot at $t=20$, $\eta\simeq0.10$, is one point on that oscillation. On the
averaged measure the inelasticity rises with multipole depth and saturates at
$\bar\eta_\infty\simeq0.156$, so the single multiplet understates the converged
value by a factor of about $2.7$: the ladder must be climbed for the physical
inelasticity to appear, and it converges rather than running away. The result is
cutoff-independent. Raising the per-mode occupation cutoff from $d=4$ to
$d=8,16,32$ moves $\bar\eta$ by at most $2\%$, and $d=8$, $16$ and $32$ coincide
at $\ell_{\max}=2,3$ where the occupation has saturated (Fig.~\ref{fig:H2}c); the
$d=8$ trajectories track the $d=4$ climb across the tower and reach a second
large register, ninety-six qubits at $\ell_{\max}=5$. Bond dimension $\chi=24$ is
likewise converged: at $\ell_{\max}=4$ the late-time inelasticity changes by
under $1\%$ between $\chi=24$ and $\chi=96$ (Appendix~\ref{app:scaling:climb}),
and the discarded weight stays below $10^{-3}$ at every register. The finite-size
caveat of the exact-diagonalization registers is thus lifted for the dynamical
and entanglement observables; it remains only for the interior level statistics,
which require many highly entangled eigenstates and lie outside the reach of the
present tensor-network evolution.

\section{Hardware resources and noise budget}
\label{sec:hardware}

\begin{figure*}[t]
\includegraphics[width=0.82\textwidth]{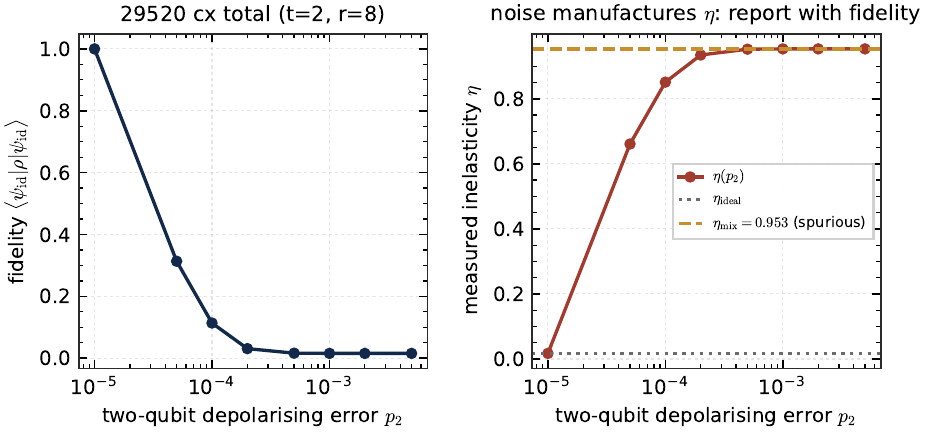}
\caption{\label{fig:G1}Depolarizing-noise study at fixed depth ($29520$ cx):
fidelity (left) and measured inelasticity (right) versus two-qubit error
rate. The spurious mixed-state value $\eta_{\rm mix}=0.953$ is marked;
the $10\%$-bias budget is $p_2=5\times10^{-5}$.}
\end{figure*}

Table~\ref{tab:resources} gives transpiled per-step costs (basis
$\{{\rm cx},R_z,\sqrt X,X\}$, optimization level 1). The depolarizing study
(Fig.~\ref{fig:G1}; 6 qubits, $t=2$, $r=8$, $29520$ cx, exact density
matrix) tracks state fidelity and the measured inelasticity \emph{jointly}:
depolarization drives $\rho$ toward the maximally mixed state, whose
spurious inelasticity $\eta_{\rm mix}=1-D_{N_0}/2^n=0.953$ dwarfs the ideal
signal $\eta_{\rm ideal}=0.0173$ at this depth. Already at
$p_2=5\times10^{-5}$ the measured $\eta=0.661$ at fidelity $0.31$; by
$p_2\ge10^{-3}$ the state is fully mixed ($F\to2^{-6}$). The
ten-percent-bias error budget is $p_2\le 5\times10^{-5}$, an order of
magnitude below contemporary two-qubit error rates at this circuit
depth, which is the quantitative basis on which hardware execution is
deferred. Noise \emph{manufactures} inelasticity; any hardware claim of
particle production must be reported jointly with fidelity. Refer to Appendix \ref{app:scaling} for detailed discussion on hardware scaling.

\begin{table}[h]
\caption{\label{tab:resources}Transpiled resources per second-order Trotter
step ($\Delta t=0.25$), Gray encoding, $d=4$.}
\begin{ruledtabular}
\begin{tabular}{lcccc}
Register & Qubits & Terms & cx & depth \\
\hline
$\ell=2$ subset (3 modes) & 6 & 283 & 3690 & 4909 \\
$\ell=2$ multiplet (5 modes) & 10 & 1103 & 12274 & 16934 \\
\end{tabular}
\end{ruledtabular}
\end{table}

\section{Discussion}
\label{sec:discussion}

\emph{Mechanism.} The results form one causal chain, and its first two links are
\emph{analytic}. The boost spectrum Eq.~\eqref{eq:omegal} is
cubic-resonance-free with gap $\to1/2$ [Eq.~\eqref{eq:gap}], and the traceless
self-coupling vanishes at leading soft order [Theorem~\ref{thm:vanish}]; these
two structural facts alone already force the leading self-interaction to be
non-thermalizing, independently of any simulation. The derived vertex magnitudes
are bounded by $0.0363\,\gt$, so at Fock-converged couplings every $\Hthree$ and
$\Hint$ process is far off shell, first-order transition amplitudes are virtual,
and the quench dresses rather than delocalizes [$d_{\rm eff}=1.06$,
Eq.~\eqref{eq:deff}]. What the simulation adds is quantitative: the measured
late-time inelasticity $\bar\eta\simeq0.06$ matches the perturbative estimate
$\sum_c(V_c/\Delta\omega)^2$; the opened four-scalar channel obeys its
fourth-order exponents exactly [Eq.~\eqref{eq:powerlaws}]; the late-time
distribution stays far from every ergodic reference; and the entanglement profile
sits $2.3$ nats under Haar. \emph{Perturbative rigidity} is the single phrase for all of it: predicted by the analytic structure, confirmed and quantified by the simulation.

\emph{Consequence for the near-horizon program.} Within the GGV framework
the $2\to2N$ tower and its Page-like entropy arise from graviton
\emph{exchange} \cite{GG2022,BGP2020}. Our result closes the alternative:
the leading soft self-interaction, treated exactly and nonperturbatively to
the full $120$-qubit multipole register, does not by itself thermalize or
scramble the longitudinal sector. Any Page-like behavior in
this framework must therefore be sourced by the exchange/eikonal dynamics,
by beyond-leading-soft kernels ($K\neq1$, whose measured systematic is
already ${\rm TV}=0.044$ at fixed structure), or by the radiative sector
outside the longitudinal truncation, a sharp narrowing of mechanism space
that was not derivable from the exchange calculations alone. The
intermediate physical-sector statistics $\rmean=0.452(16)$, rising with
$\Nmax$, mark exactly where the question ``does the self-interacting
longitudinal sector become chaotic?'' now lives.

\emph{What hardware would add.} The classical methods settle the model in its rigid, area-law regime: Aer
statevector evolution is exact at the smallest registers and the tensor-network
climb (Sec.~\ref{sec:mps}) extends that reach to the full $120$-qubit ladder, so
where the entanglement is low hardware adds no information about the model. Its
necessity begins at the crossover to volume-law entanglement; below that
crossover it still adds a demonstration layer with genuine scientific content of
its own: measured
symmetry-violation fractions as error witnesses (the exact conservation of
$L_z$, and of $Q$ with the vertex off, means every sector-violating
bitstring is a detected error, our derived Hamiltonian
provides error detection for free), benchmarking of zero-noise extrapolation, twirling, and
dynamical decoupling on a gravitationally derived Hamiltonian, and
device-sampled statistics for the protocols validated here. The budget of
Sec.~\ref{sec:hardware} defines the entry point: a reduced $d=2$ instance
of the 3--5 mode register at one or two Trotter steps
($10^2$--$10^3$ cx) with the full mitigation stack, reporting $\eta$
jointly with fidelity and the symmetry-violation fraction. Refer to Appendix \ref{app:scaling} for more discussions.

\section{Conclusion}
\label{sec:conclusion}

We derived a first-principles cubic graviton vertex on the Schwarzschild
horizon and carried it through a fully validated quantum-simulation pipeline. The
analytic structure of the vertex, a leading-soft vanishing theorem for the
traceless self-coupling together with a provably resonance-free boost spectrum
(gap $\to1/2$), \emph{predicts} that the longitudinal sector is perturbatively rigid at this order; the simulation makes it quantitative and measures the residual dressing. The confirmed statements are: exact selection rules
[$L_z$; the charge $Q=N_\phi+2N_h$; Wigner--Eckart scalarity], the resonance-free
spectrum, a structure-sensitive inelasticity (late-time average
$\bar\eta\simeq0.06$ at $\ell=2$, cutoff-converged), confirmed fourth-order power
laws $(2.000,6.000,7.92)$ for the channel that only
the vertex opens, a late-time multiplicity far from thermal/Poisson/Haar
references with $d_{\rm eff}=1.06$, an entanglement profile $2.3$ nats below
Haar, and \emph{indicative} intermediate physical-sector level statistics
$\rmean=0.452(16)$, on Hilbert spaces still too small to be decisive, made
accessible by a symmetry-exact truncation. A symmetry-graded
matrix-product-state climb of the multipole ladder to sixty modes, the
$120$-qubit register, then shows the inelasticity converging to
$\bar\eta_\infty\simeq0.16$ and the dressing remaining area-law (peak
entanglement $\simeq0.45$ nats, far below Page), establishing that the rigidity
is a property of the full register and not a finite-size artefact of the
two-multiplet system (Sec.~\ref{sec:mps}, Appendix~\ref{app:scaling:climb}).

The scope of these statements is set by the following limitations: (i)~$N$ counts occupation of the GGV longitudinal (non-radiative) sector; the map to radiative Regge--Wheeler--Zerilli
multiplicity is open. (ii)~The radial kernel is the leading soft $K=1$; the derived
polarization weight $W(\lambda_i)$ supplies the multi-$\ell$ weighting and
shifts the multi-$\ell$ longitudinal-occupation distribution by
${\rm TV}=0.093$ on $\bar p_N$ in one representative prescription
(Sec.~\ref{sec:multil}). The equal-$\ell$ weight is exact and independently
confirmed, but the off-diagonal (multi-$\ell$) weights are prescription-dependent
at the factor-of-$\sim\!3$ level (Appendix~\ref{app:theoremcheck}), so this shift
is a derived effect with a large scheme systematic.
(iii)~The background is static Schwarzschild: entropy
statements are entanglement-structure statements.
(iv)~$\gt$ sweeps the magnitude of the gravitationally fixed structure;
the physical coupling is Planck-suppressed and carries the \emph{derived}
on-shell weight $W$ (Sec.~\ref{sec:theorem}, Finding~\ref{find:cg}), so all
rates are mechanism structure; the vacuum is
the GGV in/out choice, and the odd-parity and radiative sectors, like the
$\ell=0,1$ channels, lie outside the truncation, the complete list of
modeling findings behind the derivation is collected in
Appendix~\ref{app:modeling}. (v)~Truncation systematics are
quantified where they matter: $0.014$ on $\eta$ ($d=4$ vs $6$), $21\%$ on
the absolute cascade yield ($d=3$ vs $4$; exponents robust), $4\%$ per
$\Nmax$ step on the multi-$\ell$ $\eta$. (vi)~Spectral statements are
two-sided and sector-resolved. (vii)~All quantum-circuit results are Aer
statevector/density-matrix simulations; hardware execution is deferred on
the quantitative budget of Sec.~\ref{sec:hardware}. Within that scope, the
central finding stands and now extends across the full multipole register: the
first-principles $hhh$ vertex renders the near-horizon longitudinal sector
perturbatively rigid, its inelasticity converging to
$\bar\eta_\infty\simeq0.16$ as the ladder is climbed to sixty modes, and
whatever thermalizes the black hole, it is not this vertex alone.

\begin{acknowledgments}
	I sincerely thank Dr. Diptarka Das (IIT Kanpur) for the valuable discussions and insights.
\end{acknowledgments}

\section*{Data and code availability}
Simulations presented in this paper used Qiskit 2.4.2 and Qiskit-Aer 0.17.2 with NumPy 2.4 and
SciPy 1.17; the multipole-ladder climb of Sec.~\ref{sec:mps} and
Appendix~\ref{app:scaling:climb} uses an $L_z$-graded matrix-product-state
engine built on TeNPy 1.1.0, cross-validated against the exact-diagonalization
Hamiltonian to machine precision. 
The complete simulation code (all engines), all derivation scripts (including \texttt{derive\_L3.py}, \texttt{cubic.py}, \texttt{organize.py},
\texttt{onshell.py}, \texttt{dilaton\_W.py}, \texttt{theorem\_explicit.py},
\texttt{verify\_gr.py}, \texttt{run\_verify2.py}, \texttt{verify\_euler\_identities.py}, \texttt{rerun\_derived\_kernel.py}), seeds, data, and raw outputs are available at public repository \cite{git_repo}.

\clearpage
\appendix

\section{Conventions and the explicit vertex table}
\label{app:table}

\emph{The derived on-shell weight.} The surviving trace-sector vertex of
Sec.~\ref{sec:theorem}, contracted on the GGV residue polarizations at
rest-frame kinematics ($p_i\!\cdot p_j=-\omega_i\omega_j$,
$\omega_\ell=\sqrt\lambda$, $\mu=1$; legs $1,2$ creation, leg $3$
annihilation), gives the closed-form weight $W(\lambda_1,\lambda_2,\lambda_3)$,
symmetric in $\lambda_1\leftrightarrow\lambda_2$. On the equal-$\lambda$
diagonal, $W(\lambda,\lambda,\lambda)=-3\lambda(2\lambda^2+\lambda+3)/
(\lambda+1)^2$ [Eq.~\eqref{eq:Weq}], so $W(7,7,7)=-35.4375$; this equal-$\ell$
value is exact and prescription-independent (Appendix~\ref{app:theoremcheck}).
The physical multipole weighting on a multi-$\ell$ register is carried by the
ratios $W(\lambda_i)/W_0$ with $W_0\equiv W(7,7,7)$; representative values in the
$\bdag_1\bdag_2b_3$-ordering prescription (with $\lambda_\ell=\ell^2+\ell+1$, so
$\lambda_2=7$, $\lambda_3=13$) are $W(7,7,13)/W_0=0.945$, $W(7,13,7)/W_0=2.006$,
$W(13,13,7)/W_0=3.478$, and $W(13,13,13)/W_0=3.328$, all computed with the exact
trace amplitude $\bar H=-2/(\lambda+1)$. The
asymmetry between the creation-pair legs $(1,2)$ and the annihilation leg $(3)$
reflects the operator ordering in $\Hthree$. Unlike the equal-$\ell$ value, these
off-diagonal ratios are \emph{prescription-dependent}: a different but equally
defensible off-shell continuation changes them by up to a factor of $\sim\!3$
(Appendix~\ref{app:theoremcheck}), so they set the sign and rough magnitude of
the multi-$\ell$ effect, and the
$\Delta\eta=-8.6\%$, ${\rm TV}=0.093$ of Sec.~\ref{sec:multil} inherit that
systematic.

Mode ordering on the $\ell=2$ multiplet is
$[(2,-2),(2,-1),(2,0),(2,1),(2,2)]$. Table~\ref{tab:vertex} lists all
eleven terms $V_{ij;k}\,(b_i^\dagger b_j^\dagger b_k+{\rm h.c.})$ of the
derived vertex at unit reduced coupling (with $W$ absorbed into $\gt$ on this
single-$\ell$ register); entries contain the Gaunt factor,
the phase $(-1)^{m_k}$, the normalization
$\mathcal N=(8\omega^3)^{-1/2}=0.0822$ (all $\omega=\sqrt7$), and the
distinct-pair factor. The three Wigner--Eckart witnesses of
Table~\ref{tab:validation} correspond to rows $(3,3;4)$, $(1,3;2)$, and
$(2,2;2)$ divided by $\mathcal N$ and the pair factor.

\begin{table}[h]
\caption{\label{tab:vertex}The eleven-term $\ell=2$ vertex table at
$\gt=1$ (indices in the mode ordering above; $m_i+m_j=m_k$ throughout).}
\begin{ruledtabular}
\begin{tabular}{ccr@{\hspace{1em}}ccr}
$(i,j)$ & $k$ & $V_{ij;k}$ & $(i,j)$ & $k$ & $V_{ij;k}$ \\
\hline
$(0,2)$ & $0$ & $-0.0296$ & $(1,4)$ & $3$ & $-0.0363$ \\
$(0,3)$ & $1$ & $-0.0363$ & $(2,2)$ & $2$ & $+0.0148$ \\
$(0,4)$ & $2$ & $-0.0296$ & $(2,3)$ & $3$ & $+0.0148$ \\
$(1,1)$ & $0$ & $+0.0181$ & $(2,4)$ & $4$ & $-0.0296$ \\
$(1,2)$ & $1$ & $+0.0148$ & $(3,3)$ & $4$ & $+0.0181$ \\
$(1,3)$ & $2$ & $-0.0148$ & & & \\
\end{tabular}
\end{ruledtabular}
\end{table}

\section{Validation logic of the derivation: scope of the derivation cross-check}
\label{app:validlogic}

Three analytic gates validate the derivation. (1)~\emph{Coupling
normalization}: the overall coupling of $\Vg$ must be $\gamma=\kappa/\Rs$
(Finding~\ref{find:coupling}), the same emergent coupling that controls
the matter vertex and reproduces 't~Hooft's shift coefficient
$8\pi\GN/(\ell^2+\ell+1)$ \cite{tHooft1996,DrayTHooft1985}.
(2)~\emph{Selection rules}: the Gaunt/Wigner-$3j$ rules
\eqref{eq:msel}--\eqref{eq:paritysel} are forced by the sphere integral;
any derivation violating them is wrong. (3)~\emph{Soft limit and polarization content}: as $\mu=1/R\to0$ the
curvature terms switch off and the vertex reduces to the flat-space
three-graviton vertex of DeWitt \cite{DeWitt1967} and Sannan
\cite{Sannan1986} projected onto the near-horizon polarizations. This gate
is passed by \emph{construction}: we obtain the vertex from the direct
cubic expansion of $\sqrt{-g}\,R$ (Sec.~\ref{sec:theorem}), whose flat limit
is the same object. The projection returns the vanishing theorem for the
traceless channel and the closed-form weight \eqref{eq:Weq} for the survivor;
because the derivation never averages a current, there is no unaveraged
Isaacson gauge dependence left to cancel (Sec.~\ref{sec:isaacson}).

Equally important is what the cross-check \emph{cannot} establish. The
published $2\to2N$ multiplicity, with its peak
$N_{\max}=E^2\kappa^2/e$, equal to $(2/e)\Sbh$ at $E=\Mbh$, and
multiplicity entropy $S_\Omega\sim N\log4$ \cite{GG2022,GGtoolbox}, is
produced by the effective scalar four-vertex
\begin{equation}
  V[p_1;p_2;p_3;p_4]=2\gamma^2\,p_{1a}p_{2b}\,
  \hat P^{abcd}(p_1{+}p_2)\,p_{3c}p_{4d},
  \label{eq:matterfour}
\end{equation}
i.e., by graviton \emph{exchange} through the propagator $\hat P$, with no
$h^3$ insertion; the toolbox lists graviton self-interactions among its
explicitly unstudied corrections \cite{GGtoolbox}. Therefore $H_3$ does
not enter the leading $2\to2N$ amplitude, and matching the $H_3$ tree
amplitude to the published multiplicity is not the right test, a claimed
match would recycle a solved result as an open problem. The genuinely new
prediction, the multiplicity from the cascade \eqref{eq:cascade}, has no
published counterpart to match against; that is the point, and it is why
the simulation results of Sec.~\ref{sec:results} carry discovery content
rather than reproduction.

\section{Modeling choices and approximations}
\label{app:modeling}

For transparency we gather every assumption behind
$H_3$~\eqref{eq:H3}--\eqref{eq:Vgrav}; each is a reportable finding. \textbf{(C1) Even parity only.} The odd sector decouples at
quadratic order and is suppressed at large $\ell$ (action
$\propto\lambda-1$); whether it
contributes at cubic order through mixed even--odd--odd Gaunt structures
is not settled here. \textbf{(C2) Soft truncation}
(Modeling choice~\ref{mc:soft}): transverse-derivative and
background-curvature ($\propto S_{ab}\sim\mu^2$) terms of $\tg_{ab}$ are
dropped relative to the two-derivative longitudinal terms. \textbf{(C3)
The polarization contraction, derived}
(Sec.~\ref{sec:theorem}, Finding~\ref{find:cg}): the spin-2 contraction is
fixed directly. On the traceless matter-parallel channel it vanishes at
leading soft order (Theorem~\ref{thm:vanish}); the surviving trace-sector
coupling has the closed-form weight \eqref{eq:Weq}. The residual freedoms
are two \emph{quantified} scheme sensitivities: (a)~the
trace amplitude $\bar H=-2/(\lambda+1)$ retained exactly versus its
leading-soft value $0$ ($\sim10\%$ on $W$ at $\ell=2$), and (b)~the
off-conservation-surface vertex-function prescription, of order the
off-shellness $\Delta\omega/\omega\lesssim21\%$ at $\ell=2$. \textbf{(C4)
Gauge/scheme dependence, fixed by the direct expansion}
(Sec.~\ref{sec:isaacson}):
we do not use the Brill--Hartle-averaged Isaacson current \cite{Isaacson:1968hbi,Isaacson1968,Brill:1964zz}. The vertex is
read directly off the cubic expansion of $\sqrt{-g}\,R$ in the fixed even
Regge--Wheeler gauge, so there is no averaging step to be gauge-dependent;
the only residual scheme choices are the two quantified sensitivities of
(C3). \textbf{(C5) Trace/$K$ decoupling, quantified, not assumed}: the
theorem shows the traceless self-coupling vanishes and the physical coupling
runs precisely \emph{through} the trace/$K$ sector, so trace/$K$ do not
decouple at cubic order, as anticipated. The multi-$\ell$ effect of the
derived $\lambda$-dependence is measured directly in Sec.~\ref{sec:multil}
(a $-8.6\%$ shift in the longitudinal-occupation inelasticity, total-variation
distance $0.093$).
\textbf{(C6) The $\ell=0,1$ modes} (Modeling choice~\ref{mc:low}):
gauge-special, with no radiative content and pathological pole residues;
our registers use $\ell\ge2$. \textbf{(C7) Faddeev--Popov ghosts}:
non-propagating for $\ell\ge2$, instantaneous for $\ell=0,1$
\cite{KalloshRahman2021}; at tree level for the Hamiltonian they do not
contribute, stated rather than assumed away. \textbf{(C8) Vacuum
dependence}: the number operator, and hence the inelasticity observable,
depends on the Boulware/Unruh/Hartle--Hawking choice; we adopt the GGV
in/out basis. \textbf{(C9) Hierarchy truncation}
(Sec.~\ref{sec:secondquant}): the maximally off-shell $\Delta N=\pm3$ pieces are
dropped in favour of the least off-shell $\Delta N=\pm1$ pieces; this is a
truncation ordered by off-shellness (the
spectrum is resonance-free, so no retained term is on shell), and
Sec.~\ref{sec:selection}(c) quantifies the parametric separation that justifies
it. \textbf{(C10) Non-radiative vs.\ radiative graviton}:
$H_3$ is the self-coupling of the GGV longitudinal shock mode, distinct
from the radiative Regge--Wheeler--Zerilli graviton
\cite{MartelPoisson2005,KalloshRahman2021}; $N$ counts occupation of the
longitudinal sector, a well-defined simulation observable whose map to a
physical radiated-graviton multiplicity, and with it the reconciliation
of the GGV in/out basis with the Kallosh--Rahman mode functions, is
open. If the complete spin-2 vertex proves intractable, the object
\eqref{eq:H3}--\eqref{eq:Vgrav}, the dominant, gravity-fixed, soft,
angular-momentum-conserving vertex with exact $\gamma$ and exact Gaunt
structure, remains a defensible and vastly more physical Hamiltonian
than any hand-chosen $V_{ijk}$; stating precisely what is retained is
itself a result.

\section{Numerical verification of the expansion coefficients}
\label{app:numcheck}

The two coefficients on which the derivation pivots, the $-\tfrac14$ of
$S^{(2)}$ in \eqref{eq:S2} and the $\kappa^3/6$ of
\eqref{eq:euler3}--\eqref{eq:S3}, were verified numerically, in the
two-engine spirit of this paper, by an exact-nonlinear-GR spectral
computation independent of any perturbative expansion. On a periodic
$16^4$ lattice with Fourier derivatives, draw a random smooth symmetric
perturbation $h_{\mu\nu}(x)$ built from low wavevectors closed under
zero-sum triads ($k_a+k_b+k_c=0$), so the triple products entering the
cubic invariant stay below the Nyquist frequency. Form the exact metric $g=g^0+\kappa h$ at several $\kappa$, compute the
exact nonlinear Christoffel, Riemann, and Einstein tensors, and the two
box-averaged scalars
\begin{equation}
  F(\kappa)=\big\langle\sqrt{-g}\,R\big\rangle,\qquad
  J(\kappa)=\big\langle h_{\mu\nu}\sqrt{-g}\,G^{\mu\nu}\big\rangle
  \nonumber
\end{equation}
(full-metric raising, measure included). Box-averaging annihilates every
spectral total derivative exactly, so Euler homogeneity of
$\int\sqrt{-g}R$ implies the functional identity $\dd F/\dd\kappa=-J$
order by order; the Taylor coefficients $\{s_n\}$ of $F$ and $\{j_n\}$ of
$J$ are extracted by an even/odd split with Richardson extrapolation
(well-conditioned, unlike a direct multi-term fit whose near-collinear
columns $\kappa^2,\kappa^3,\dots$ corrupt the subleading coefficient).
Four diagnostics, on four independent random seeds:
(1)~\emph{pipeline consistency}: the finite-difference
$\dd F/\dd\kappa+J$ vanishes to $\lesssim10^{-12}$ at $\kappa=0.03$,
confirming that the independently coded nonlinear $R$ and $G^{\mu\nu}$
pipelines agree to machine precision;
(2)~\emph{quadratic coefficient}:
$c_2=S^{(2)}/\langle hG^{(1)}\rangle=(s_2/2)/j_1=-0.250000$ on every
seed, confirming the $-\tfrac14$ of \eqref{eq:S2} in our convention (the
toolbox's printed $-\tfrac12$ corresponds to a different field
normalization \cite{GGtoolbox});
(3)~\emph{Euler identities}: $2s_2/j_1=-1.000000000$ and
$3s_3/j_2=-1.000000000$ on every seed, the content of
$n\,s_n=-j_{n-1}$ at $n=2,3$, with the cubic case verified to ten
significant figures, fixing the $\kappa^3/6$ of \eqref{eq:S3};
(4)~\emph{the discriminating negative control}: the naive contraction
$\langle h_{\mu\nu}G^{(2)\mu\nu}\rangle$ with $\eta$-raised indices and
no measure gives a seed-dependent, non-universal ratio (e.g.\ $-0.14$,
$-0.10$, $-0.90$, $+0.23$ across the four seeds), the exact identity
holds only for the densitized, full-metric-raised current, exactly as
required by the discussion below Eq.~\eqref{eq:S3}.

\subsection{Verification of the vanishing theorem and the derived weight}
\label{app:theoremcheck}

The direct cubic expansion of Sec.~\ref{sec:theorem} was carried out and
cross-checked by three independent engines. \emph{Engine~A (symbolic
Lagrangian).} Inserting the two-block near-horizon metric into
$\sqrt{-g}\,R$ and collecting $O(\kappa^3)$ yields the cubic Lagrangian
$\mathcal L_3$ ($76$ differential monomials); its momentum-space vertex,
$375$ multilinear terms before reduction, is reduced on the conservation
surface $p_3=-p_1-p_2$. Setting the polarizations purely traceless with the
trace/transverse scalar $K=0$ collapses the reduced vertex to $0$
\emph{identically} (exact symbolic zero), which is Theorem~\ref{thm:vanish}.
\emph{Engine~B (exact nonlinear GR).} On a $16^2$ spectral grid the
$\kappa^3$ Taylor coefficient of $\langle\sqrt{-g}\,R\rangle$, extracted by
the same even/odd Richardson procedure, is at machine zero
($|s_3|\sim10^{-13}$--$10^{-10}$) on four random traceless seeds, while the
generic (non-traceless) seeds reproduce the symbolic $\mathcal L_2,\mathcal
L_3$ to $\sim10^{-6}$ relative (Richardson-limited). \emph{Engine~C (fully
symbolic GR cross-check).} An independent closed-form Christoffel/Ricci
computation on a fixed zero-sum triad returns the $\kappa^3$ mean $+1/4$;
this engine caught and fixed a sign error in the inverse-metric Neumann
ordering of Engine~A (the corrected expansion reproduces $+1/4$), so the
surviving-sector coefficients are sign-correct. The surviving vertex was
then contracted on the GGV residue polarizations (traceless amplitude
$P=\lambda/(\lambda+1)$, $K=-1$, trace $\bar H=-2/(\lambda+1)$) at
rest-frame kinematics, giving the closed form \eqref{eq:Weq},
$W(7,7,7)=-2268/64=-35.4375$, symmetric under exchange of the two creation
legs to machine precision.

\emph{Engine~D (independent dilaton-form derivation).} A third symbolic engine
catching a sign error in the first is reassuring, but it also motivates an
analytic check that does not share the substrate of Engines~A/C; we therefore
re-derived the weight by an independent route that never forms the $375$-term
four-dimensional vertex. Starting from the exact
warped-product identity \eqref{eq:dilatonid}, the cubic Lagrangian of the
surviving sector is read off in closed form as $\mathcal L_3=K\,[\sqrt{-g^{(2)}}
R^{(2)}]^{(2)}+\tfrac12[\sqrt{-g^{(2)}}g^{(2)ab}]^{(1)}\partial_aK\partial_bK
-\tfrac12 K\,\eta^{ab}\partial_aK\partial_bK$ (nineteen monomials, versus
seventy-six by direct expansion), whose momentum-space contraction on the same
GGV residue polarizations gives
\begin{equation}
\begin{aligned}
  W(\lambda,\lambda,\lambda)&=-\frac{3\lambda(2\lambda^2+\lambda+3)}{(\lambda+1)^2},\\
  W(7,7,7)&=-\frac{567}{16}=-35.4375,
\end{aligned}
  \nonumber
\end{equation}
\emph{identical} to Engines~A/C on the entire equal-$\lambda$ diagonal (ratio
$1$ symbolically, all $\lambda$). This is an analytic check: it
reproduces the boxed weight from a manifestly different Lagrangian and confirms
the $\kappa^3/6$ coefficient, retiring the sign-error worry. The same route also confirms the sharper Remark~\ref{rem:strong}: setting
$K=0$ with $\bar H\neq0$ annihilates the reduced vertex identically. Script
\texttt{dilaton\_W.py} is openly available at \cite{git_repo}.

\emph{Prescription-dependence of the off-diagonal weights.} The equal-$\lambda$
weight is prescription-\emph{independent}: at coincident $\lambda_i$ the boost
frequencies coincide and the conservation-surface, all-on-shell, and mixed
prescriptions all collapse to the same pure-polarization contraction, which is
why Engine~D and Engines~A/C agree there to machine precision. The off-diagonal
weights $W(\lambda_1,\lambda_2,\lambda_3)$ enjoy no such protection. Evaluating
Engine~D under four defensible off-shell prescriptions---(H0) the conservation
surface $\omega_3=\omega_1+\omega_2$ with $\omega_{1,2}=\sqrt{\lambda_{1,2}}$;
(H1) all three legs on shell, $\omega_i=\sqrt{\lambda_i}$, conservation not
imposed; (H2) the sign convention $\omega_3=+(\omega_1+\omega_2)$; and
(H3) $\omega_3=-\sqrt{\lambda_3}$, produces ratios $W(\lambda_i)/W(7,7,7)$ that
differ substantially, e.g. for $(\ell_1,\ell_2,\ell_3)=(2,3,2)$ (so
$\lambda=7,13,7$) the ratio spans $0.95$ (H3) to $1.46$ (H0) to $2.01$ (paper's
Appendix~\ref{app:table} value), and for $(3,3,2)$ it spans $0.83$ to $3.48$, a
factor of $\sim\!3$--$4$ across prescriptions, and up to $\sim\!70\%$ between our
most defensible on-conservation-surface value (H0) and the Appendix
value on the largest entries. Only H0 (and, up to an overall sign, H3) returns a
diagonal $W(7,7,7)$ consistent with the boxed $-35.4375$; H1 and H2 rescale the
diagonal, confirming that the off-diagonal continuation is genuinely a scheme
choice rather than a fixed kinematic fact. The Appendix~\ref{app:table}
ratios coincide with none of H0--H3 exactly, which is itself a measure of the
prescription latitude. We therefore do not treat the off-diagonal ratios as sharp
numbers. The single-$\ell$ renormalization is exact; the multi-$\ell$ weighting is
a derived \emph{effect} carrying this large scheme systematic, and the
$\Delta\eta=-8.6\%$, ${\rm TV}=0.093$ of Sec.~\ref{sec:multil} is the value in one
representative prescription, reported with that caveat rather than as a fixed
prediction. Propagating any single prescription through the simulation model
reproduces the $\ell=2$ tables as an exact constant multiple ($-35.4375\times$
the $K=1$ table, constant to $10^{-12}$), since on a single multiplet the
off-diagonal freedom does not enter.

All scripts (\texttt{derive\_L3.py}, \texttt{cubic.py}, \texttt{organize.py},
\texttt{onshell.py}, \texttt{dilaton\_W.py}, \texttt{theorem\_explicit.py},
\texttt{verify\_gr.py}, \texttt{rerun\_derived\_kernel.py}) are available at \cite{git_repo}.

\section{Explicit order-by-order collapse of the traceless self-coupling}
\label{app:explicit}

It is worth exhibiting the cancellation of Theorem~\ref{thm:vanish} term by
term, rather than only inferring it from Gauss--Bonnet; we give it here in full. By the
reduction of Remark~\ref{rem:strong} and Eq.~\eqref{eq:dilatonid}, the
two-dimensional block carries the entire content of the theorem: with
$\bar H=K=0$ the transverse sphere is flat and undeformed, so
$\sqrt{-g^{(4)}}R^{(4)}=\Omega^2\sqrt{-g^{(2)}}R^{(2)}\big|_{\Omega=1}$ equals
$\sqrt{-g^{(2)}}R^{(2)}$ up to the constant sphere volume, and the
four-dimensional cubic self-coupling equals the two-dimensional one.

\emph{Parametrization.} In the $(t,z)$ block with
$\eta_{ab}=\mathrm{diag}(-1,+1)$, tracelessness
$\eta^{ab}h_{ab}=h_{zz}-h_{tt}=0$ forces $h_{tt}=h_{zz}\equiv u$, leaving the
off-diagonal component $h_{tz}\equiv v$, so that
\begin{equation}
  g^{(2)}_{ab}=\begin{pmatrix}-1+\kappa u & \kappa v\\[2pt]
  \kappa v & 1+\kappa u\end{pmatrix},
  \qquad u,v=u(t,z),\,v(t,z),
  \label{eq:g2traceless}
\end{equation}
with $\det g^{(2)}=-\big[1-\kappa^2(u^2-v^2)\big]$. Here $u$ and $v$ are the two
independent traceless longitudinal polarizations.

\emph{Order-by-order reduction.} Expanding the exact scalar density
$\mathcal L\equiv\sqrt{-g^{(2)}}R^{(2)}$ in $\kappa$, each order is a manifest
total derivative:
\begin{align}
  \mathcal L^{(1)}&=\partial_t(\partial_t u-\partial_z v)
     +\partial_z(\partial_z u-\partial_t v),\label{eq:L1coll}\\[2pt]
  \mathcal L^{(2)}&=-\big[\partial_t(u\,\partial_z v)-\partial_z(u\,\partial_t v)\big]
     =-(\partial_t u\,\partial_z v-\partial_z u\,\partial_t v),\label{eq:L2coll}\\[2pt]
  \mathcal L^{(3)}&=\partial_t\!\Big[\tfrac12(u^2-v^2)(\partial_t u-\partial_z v)\Big]\nonumber\\
     &\quad+\partial_z\!\Big[\tfrac12(u^2-v^2)(\partial_z u-\partial_t v)\Big].\label{eq:L3coll}
\end{align}
The linear term \eqref{eq:L1coll} is the linearized Ricci density, a total
derivative by construction. The quadratic term \eqref{eq:L2coll} is the Jacobian
$-\,\partial(u,v)/\partial(t,z)$, a null Lagrangian (a total derivative for any
$u,v$).

\emph{The cubic term, before organizing.} The object whose bulk part would be
the graviton self-vertex is the thirteen-monomial expression
\begin{equation}
\begin{aligned}
  \mathcal L^{(3)}={}& u\,(\partial_t u)^2+u\,(\partial_z u)^2\\
   &+\tfrac12 (u^2-v^2)\big(\partial_t^2 u+\partial_z^2 u\big)\\
   &+(v^2-u^2)\,\partial_t\partial_z v
    +2v\,\partial_t v\,\partial_z v\\
   &-v\big(\partial_t u\,\partial_t v+\partial_z u\,\partial_z v\big)\\
   &-u\big(\partial_t u\,\partial_z v+\partial_z u\,\partial_t v\big),
\end{aligned}
  \label{eq:L3raw}
\end{equation}
and it collapses \emph{exactly} to the divergence \eqref{eq:L3coll}, with no
remainder. Equivalently, and most physically, the cubic vertex is the
Euler--Lagrange derivative of $\int\mathcal L^{(3)}$, and it vanishes identically
in both polarizations,
\begin{equation}
  \frac{\delta}{\delta u}\!\int\!\mathcal L^{(3)}\,d^2x
  =\frac{\delta}{\delta v}\!\int\!\mathcal L^{(3)}\,d^2x\;\equiv\;0,
  \label{eq:ELzero}
\end{equation}
so there is no bulk cubic interaction, on shell or off, of the traceless
longitudinal mode with itself. The same vanishing holds at quadratic order
[\eqref{eq:L2coll} is a null Lagrangian, so its Euler--Lagrange derivative is also
zero], which means the traceless mode has no kinetic term \emph{from the
two-dimensional Einstein--Hilbert density alone}: its dynamics come entirely from
the transverse coupling switched on by $K\neq0$, which is precisely the surviving
trace-sector vertex of Sec.~\ref{sec:theorem}.

\emph{Unifying statement.} Equations~\eqref{eq:L1coll}--\eqref{eq:L3coll} are the
$O(\kappa^{1,2,3})$ instances of Gauss--Bonnet: because
$\int d^2x\,\sqrt{-g^{(2)}}R^{(2)}=4\pi\chi$ is a topological invariant, its
variation to every order in $\kappa$ is a boundary term, and the potentials above
exhibit those boundary terms explicitly. This is the term-by-term cancellation as verified symbolically in \texttt{theorem\_explicit.py} \cite{git_repo},
which also confirms \eqref{eq:ELzero} directly.

\section{Elastic-benchmark calibration of the circuit machinery}
\label{app:elas-bench}

\begin{figure}[h]
\includegraphics[width=\columnwidth]{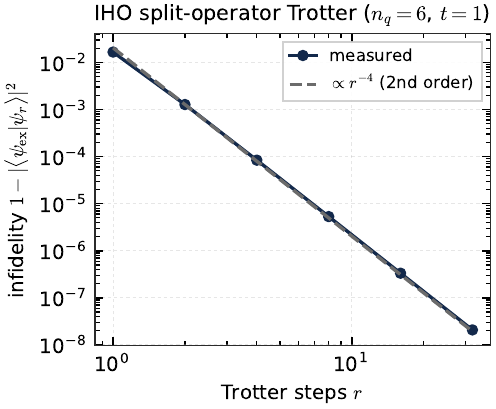}
\caption{\label{fig:A1}Elastic-benchmark Trotter-error calibration: genuine $r^{-4}$
infidelity scaling of the decomposed second-order circuits.}
\end{figure}

\begin{figure}[h]
\includegraphics[width=\columnwidth]{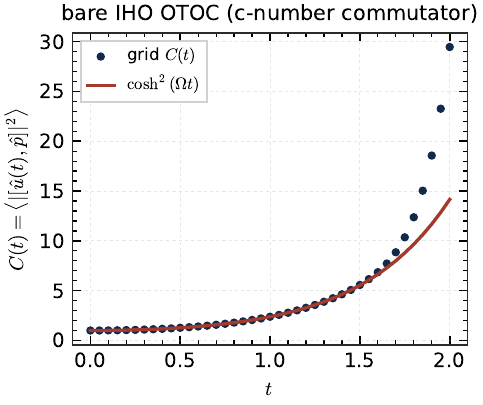}
\caption{\label{fig:A2}Inverted-oscillator squared commutator versus the analytic
$\cosh^2\Omega t$ (relative deviation $1.7\times10^{-10}$ in the validity
window).}
\end{figure}

The elastic backbone of the near-horizon S-matrix, the inverted harmonic
oscillator of the boost Hamiltonian and the Dray--'t~Hooft shift
map \cite{tHooft1996,DrayTHooft1985}, is exactly solvable and serves as the instrument calibration.
On a symmetric position grid with centered-QFT momentum representation:
(i)~Strang-split evolution shows the genuine second-order global scaling,
infidelity $1.65\times10^{-2}\to2.08\times10^{-8}$ for $r=1\to32$ at fixed
$t$ (consecutive ratios $\to16$; fitted slope $-4.03$)
[Fig.~\ref{fig:A1}]; (ii)~the grid squared commutator reproduces
$\cosh^2\Omega t$ to relative $1.7\times10^{-10}$ before wrap-around
[Fig.~\ref{fig:A2}]; (iii)~the cross-register diagonal implementation of
$S=\exp(ic\,p_{\rm in}p_{\rm out})$ is exact to $5.0\times10^{-15}$;
(iv)~the IHO spectrum in the box is regular ($\rmean=0.731$,
level-clustered), as expected for an integrable system.

\section{Measurement protocols}
\label{app:protocols}

\emph{Segment-snapshot evolution.} The transpiled second-order segment
$U_T(\Delta t)$ ($r$ repetitions) is applied iteratively on the Aer
statevector backend, saving the state after each segment and injecting it
as the next initial state; memory is bounded by one segment and the
recorded dynamics is exactly the Trotterized one, with error accumulating
linearly in the number of segments as on hardware. Production values:
$\Delta t=2$, $r=16$ per segment (twenty segments to $t=40$), giving the
$1.6\times10^{-3}$ agreement of Sec.~\ref{sec:onset}.

\emph{Multiplicity readout.} Computational-basis samples are decoded
bitstring$\to$occupations$\to N$; estimates carry binomial standard errors.
The $t=20$ sampling circuit uses $r=60$ total ($\sim10^{-2}$ Trotter
level), $16384$ shots.

\emph{Squared commutator.} With unitary Pauli probes $W,V$ the MSS
squared commutator is $C=2-2\,{\rm Re}\,
\langle\psi|W^\dagger(t)V^\dagger W(t)V|\psi\rangle$, measured as the
ancilla $\langle Z\rangle$ of a Hadamard test on the controlled sequence;
controlled time evolution is avoided by conjugation,
${\rm c}\!-\!(U^\dagger VU)=U^\dagger\,{\rm c}V\,U$, so only the probes are
controlled. Validation: ten time points, $16384$ shots, maximal absolute
deviation $0.102$ from the dense engine on $C\in[0,1.37]$, Trotter-dominated
[Fig.~\ref{fig:E1}(b)].

\section{Scaling beyond the simulable multiplet: register size, hardware budget, and the accessible physics}
\label{app:scaling}

The dynamical results of the main text are obtained on registers spanning one
or two angular-momentum multiplets. This appendix quantifies the size of those
registers and the origin of their limitations, tabulates the qubit and
gate cost of enlarging them, delimits the hardware regime such enlargement
would require, and states the physical questions that only larger systems can
settle.

\subsection{The size of the simulated system and the origin of its limitations}
\label{app:scaling:size}

The production simulations use the $\ell=2$ multiplet (five modes, ten qubits)
and the $\ell=2\oplus3$ register (twelve modes), with a per-mode occupation
cutoff $d=4$ for production and $d=5,6$ as convergence overlays. Two independent
quantities set the size of such a register, and they constrain the physics in
very different ways.

The per-mode cutoff $d$ is a convergence control. The observables are already converged in it: the inelasticity varies by
$|\eta_{d=4}-\eta_{d=6}|\le0.0144$ across the entire coupling sweep, and the
late-time multiplicity distribution shifts by a comparable amount. Raising $d$
to $8$ or $16$ would reproduce the same numbers at greater cost, without
altering any conclusion; the cutoff overlays are evidence that the truncation is safe.

The quantity that actually limits the physics is the number of modes, and with
it the dimension of the accessible Hilbert space. On one or two multiplets this
dimension is small, with two direct consequences. First, the effective
dimension of the initial state over energy eigenstates is $d_{\mathrm{eff}}=
(\sum_n|c_n|^4)^{-1}=1.06$: the quench dresses a single eigenstate rather than
delocalizing, and the energy shell $E_0\pm\sigma_E$ contains one eigenstate at
the widths explored. Second, under the symmetry-exact total-occupation
truncation the physical fixed-$(L^2,L_z)$ sectors remain of dimension
$18$--$72$ even at $\Nmax=7$, far below the size at which a level-spacing
distribution or an out-of-time-order growth window attains its asymptotic form;
at $d_{\mathrm{eff}}\simeq1$ the squared commutator is bounded and non-growing
almost by construction.

The consequence for interpretation is the central limitation of the study, and
it is a statement about size rather than about the value of any single cutoff.
At these dimensions the perturbative rigidity that the simulation measures
cannot be cleanly separated from finite-size saturation: a bounded squared
commutator and intermediate level statistics are generic signatures of a small
Hilbert space as much as of a genuinely non-scrambling dynamics. The rigidity
does possess an analytic origin independent of system size, the resonance-free
boost spectrum, whose smallest allowed cubic mismatch is bounded below by
$0.565/\Rs$ and tends to $1/(2\Rs)$, together with the leading-soft vanishing
theorem for the traceless self-coupling, but a decisive separation of the two
readings requires the larger registers quantified below. The sector-averaged
level-spacing ratio rising from $\rmean=0.351$ at $\Nmax=4$ to $0.452(16)$ at
$\Nmax=5$ is the sharpest indication that the separation is worth making: it is
the behaviour that an emergent chaos setting in with occupation, or alternatively
a receding finite-size artefact, would each produce at the accessible depths.

\subsection{Qubit and gate budgets for larger registers}
\label{app:scaling:budget}

A register spanning the multipoles $\ell=2,\dots,\ell_{\max}$ contains
\begin{equation}
	M(\ell_{\max})=\sum_{\ell=2}^{\ell_{\max}}(2\ell+1)
	=(\ell_{\max}+1)^2-4
	\label{eq:modecount}
\end{equation}
angular-momentum modes, each mapped to $n_b=\lceil\log_2 d\rceil$ qubits, so the
register occupies $M\,n_b$ qubits. The number of independent cubic vertices is
fixed by the selection rules of the derived Hamiltonian: a term
$\bdag_i\bdag_j b_k$ is nonzero only for $m_i+m_j=m_k$, the triangle inequality
on $(\ell_i,\ell_j,\ell_k)$, and even $\ell_i+\ell_j+\ell_k$. Counting the
distinct such terms with the creation pair unordered reproduces the
eleven-term $\ell=2$ vertex table of the main text exactly, which fixes the
counting convention; the vertex counts in Table~\ref{tab:scaling} are the
exact values returned by the assembled Hamiltonian.

The number of Pauli strings in the qubit Hamiltonian is dominated by the cubic
term and scales, at fixed encoding, linearly with the vertex count. The $\ell=2$
multiplet's $1103$-term \texttt{SparsePauliOp} corresponds to its eleven
vertices at $\approx\!100$ strings per vertex in the Gray $d=4$ encoding, the
diagonal quadratic terms contributing negligibly, and the transpiled
two-qubit-gate count per second-order Trotter step is $\approx\!11$ times the
string count. Applying these two calibrations, taken from the main-text
resource tables, gives the estimates in Table~\ref{tab:scaling}. The mode,
qubit, and vertex counts there are exact; the string and gate counts are
calibrated estimates that overcount mildly at large $\ell_{\max}$, where
distinct vertices increasingly share Pauli strings, and whose exact values
require assembling the operator. The $\ell_{\max}=7$ row is the $110$--$120$
qubit register: $60$ modes, $120$ qubits at $d=4$, $3454$ gravity-allowed
cubic vertices, about $3.5\times10^{5}$ Pauli strings, and about
$3.9\times10^{6}$ two-qubit gates per Trotter step, against the ten qubits and
$1103$ strings of the multiplet simulated here. This is the register that the
tensor-network climb of Appendix~\ref{app:scaling:climb} reaches classically.

\begin{table*}[t]
	\caption{\label{tab:scaling}Register-scaling budget for the graviton
		self-interaction Hamiltonian $H_2^{\rm grav}+\Hthree$. The mode count
		$M=(\ell_{\max}+1)^2-4$, the qubit counts at per-mode cutoff $d=4$ ($n_b=2$)
		and $d=8$ ($n_b=3$), and the number of gravity-allowed cubic vertices are
		exact; the counting reproduces the eleven-term $\ell=2$ vertex table of the
		main text. Pauli-string counts (Gray encoding, $d=4$) and transpiled
		two-qubit-gate counts per second-order Trotter step are calibrated to the
		measured $\ell=2$ build ($11$ vertices $\to1103$ strings $\to12{,}274$ gates)
		and are order-of-magnitude estimates. Adding the matter sector for the
		$\phi\phi\to hh\to4\phi$ cascade raises the counts further. The $\ell_{\max}=7$
		row is the $110$--$120$ qubit target.}
	\begin{ruledtabular}
		\begin{tabular}{ccccccc}
			$\ell_{\max}$ & modes $M$ & qubits ($d{=}4$) & qubits ($d{=}8$)
			& cubic vertices & Pauli strings$^{a}$ & 2-qubit gates / step$^{a}$ \\
			\hline
			$2$ & $5$   & $10$  & $15$  & $11$    & $1.1\times10^{3}$ & $1.2\times10^{4}$ \\
			$3$ & $12$  & $24$  & $36$  & $53$    & $5.3\times10^{3}$ & $5.9\times10^{4}$ \\
			$4$ & $21$  & $42$  & $63$  & $253$   & $2.5\times10^{4}$ & $2.8\times10^{5}$ \\
			$5$ & $32$  & $64$  & $96$  & $713$   & $7.2\times10^{4}$ & $8.0\times10^{5}$ \\
			$6$ & $45$  & $90$  & $135$ & $1690$  & $1.7\times10^{5}$ & $1.9\times10^{6}$ \\
			$7$ & $60$  & $120$ & $180$ & $3454$  & $3.5\times10^{5}$ & $3.9\times10^{6}$ \\
			$8$ & $77$  & $154$ & $231$ & $6438$  & $6.5\times10^{5}$ & $7.2\times10^{6}$ \\
		\end{tabular}
	\end{ruledtabular}
	\footnotesize $^{a}$Calibrated estimate; the $\ell=2$ row reproduces the
	measured $1103$ Pauli strings and $12{,}274$ transpiled two-qubit gates.
\end{table*}

\subsection{Feasibility of large-register hardware execution}
\label{app:scaling:hardware}

Enlarging the register raises two separate hardware questions with opposite
answers: whether the qubits can be accommodated, and whether the resulting
circuit can be executed with useful fidelity.

Accommodation is not a constraint. Contemporary superconducting processors
provide of order $10^{2}$ physical qubits on a single device, heavy-hexagonal
tunable-coupler chips at $133$ and $156$ qubits, a $120$-qubit square-lattice
device, and a $1121$-qubit testbed \cite{IBMroadmap}, so the $120$-qubit
$\ell\le7$ register of Table~\ref{tab:scaling} fits comfortably on present
hardware.

Execution is the binding constraint, because these devices are depth-limited
rather than width-limited. Current hardware executes of order a few thousand
two-qubit gates before noise overwhelms the signal; utility-scale
demonstrations reach $\sim\!3\times10^{3}$--$5\times10^{3}$ two-qubit gates at a
per-gate error near $10^{-3}$ \cite{Kim2023,IBMroadmap}. The depolarizing-noise
analysis of the main text sets the error rate this model requires: at ten
qubits a ten-percent bias on the inelasticity demands a two-qubit error
$p_2\lesssim5\times10^{-5}$, and by $p_2\gtrsim10^{-3}$ the state is fully mixed,
its spurious inelasticity $\eta_{\mathrm{mix}}\simeq0.95$ dwarfing the ideal
signal. Even the smallest register therefore sits at the edge of what present
hardware can do, and noise manufactures apparent particle production that must
be reported jointly with fidelity.

The larger registers lie far outside this budget. The $\ell\le7$ register
requires $\sim\!3.9\times10^{6}$ two-qubit gates per Trotter step
(Table~\ref{tab:scaling}); keeping the expected number of gate errors below
unity within a single step demands $p_2\lesssim2.5\times10^{-7}$, roughly four
orders of magnitude below the best current two-qubit gates, and the single-step
gate count already exceeds the projected near-term circuit-depth capacity
($\sim\!10^{4}$ gates) by more than two orders of magnitude, before the
$\mathcal{O}(10)$ Trotter steps and $\mathcal{O}(10)$ time points of a
dynamical trajectory are accounted for. Meaningful execution of the full
multi-$\ell$ Hamiltonian at $110$--$120$ qubits is thus a fault-tolerant task,
requiring error-corrected logical qubits rather than the noisy physical qubits
of the present generation.

A more immediately useful route is indicated by the entanglement structure the
simulation itself reports. The produced state is weakly entangled, a full
$2.3$ nats below the Haar (Page) value at the half-cut, a structured dressing
cloud rather than a random state. Low bipartite entanglement is precisely the
regime in which classical tensor-network methods, matrix-product-state time
evolution and the density-matrix renormalization group
\cite{White1992,Schollwock2011}, are efficient and can reach mode numbers far
beyond the reach of a noisy quantum device. For a perturbatively rigid,
low-entanglement sector these methods are the natural instrument for climbing
the multipole ladder and testing whether the rigidity persists as the register
grows; that climb is carried out in Appendix~\ref{app:scaling:climb}. Quantum hardware becomes necessary, and advantageous, only in the
regime where the entanglement grows toward the Page value, i.e.\ where the
dynamics begin to scramble and the matrix-product representation breaks down;
that crossover, rather than the qubit count alone, is the physically motivated
trigger for hardware execution.

\subsection{The tensor-network climb: method, validation, and results}
\label{app:scaling:climb}

The route indicated above has been carried out. This subsection records the
tensor-network engine, its cross-validation against the exact-diagonalization
Hamiltonian, and the results of climbing the multipole ladder to the
$120$-qubit register.

\paragraph{Representation and conserved charge.}
The Hamiltonian $H_2^{\rm grav}+\Hthree$ is represented as a matrix-product
operator over the $M(\ell_{\max})$ modes. The cubic vertex changes the graviton
number by one and the matter charge $Q$ by two, so no total occupation is
available as a grading; the only exactly conserved abelian quantum number is
the azimuthal weight $L_z=\sum_i m_i n_i$. Assigning each bosonic mode the
$U(1)$ charge $m_i n_i$ makes the operator $\bdag_i\bdag_j b_k$ carry net charge
$m_i+m_j-m_k$, which the matrix-product operator admits only when it vanishes;
the Gaunt selection rule $m_i+m_j=m_k$ is thereby enforced by the symmetry
structure itself, and the single-$(2,0)$ quench evolves within the fixed
$L_z=0$ sector. The vertex coefficients are the same $V_{ijk}$ used in the
statevector engine.

\paragraph{Evolution.}
Two features of $\Hthree$ dictate the integrator. It is a genuinely long-range
three-body vertex, coupling non-adjacent modes through the Gaunt rule, so
two-site time-dependent-variational-principle (TDVP) evolution \cite{Haegeman2016} cannot generate
the entangling channel from the product initial state: the local two-site
projector has no matrix element to a split across non-adjacent sites, the known
long-range TDVP cold-start. Matrix-product-operator evolution of the
Zaletel~\emph{et~al.}\ $W^{\rm II}$ form \cite{Zaletel2015} does seed the
entanglement but costs $\mathcal{O}((\chi\,\chi_{\rm MPO})^2)$ per step, with
the operator bond dimension $\chi_{\rm MPO}$ growing along the ladder
($41,114,228,429$ at $\ell_{\max}=3,4,5,6$). We therefore evolve by a short
$W^{\rm II}$ warm-up at a tight bond cap, purely to move the state off the
product manifold, followed by two-site TDVP, whose cost
$\mathcal{O}(\chi^2\,\chi_{\rm MPO})$ is linear in $\chi_{\rm MPO}$, for the
remainder. Modes are ordered so that $\pm m$ partners are adjacent, localizing
the dominant $(m,-m)$ splitting channel and minimizing the bond dimension, the
tensor-network counterpart of orbital ordering in quantum-chemistry DMRG. The
total-quanta distribution $P(N,t)$ is obtained from the characteristic function
$\langle e^{i\theta N}\rangle$, a bond-dimension-one product operator, by an
inverse discrete Fourier transform, so no $d^{M}$ statevector is ever formed.

\paragraph{Validation.}
On the $\ell=2$ multiplet, where the per-mode-truncated Hilbert space is the
same in both engines, the matrix-product-operator spectrum in the $L_z=0$
sector agrees with the exact-diagonalization Hamiltonian to $6.8\times10^{-14}$
($d=4$, $112$ levels) and $1.2\times10^{-13}$ ($d=6$, $552$ levels), validating
every matrix element including the coincident-index terms. The real-time
observables agree with exact block evolution to $\sim\!10^{-3}$, the
integrator-limited tolerance, matching the Trotter accuracy of the statevector
engine.

\begin{figure}[t]
\includegraphics[width=\columnwidth]{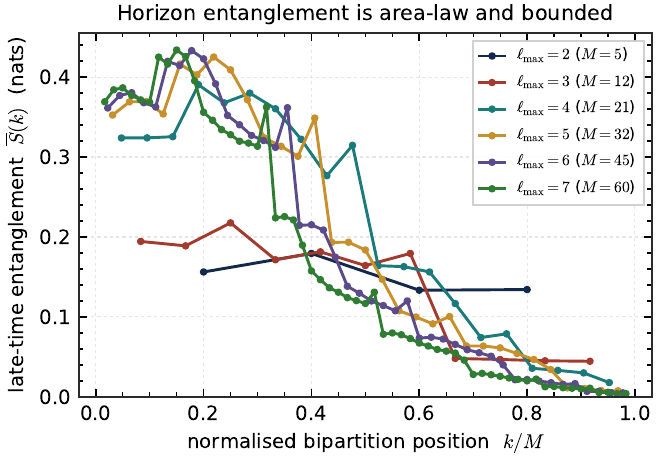}
\caption{\label{fig:H1}Late-time-averaged mode-bipartition entanglement profile
$\overline{S}(k)$ across the ordered mode chain for each multipole depth. The
entanglement peaks near the active low-$|m|$ modes and decays; the
$\ell_{\max}=4$--$7$ profiles collapse onto a common envelope of peak
$\simeq0.45$ nats, an area-law dressing that has converged in $M$. The small
even/odd sawtooth reflects bonds cutting between a $\pm m$ partner pair versus
between pairs.}
\end{figure}

\begin{figure}[t]
\includegraphics[width=\columnwidth]{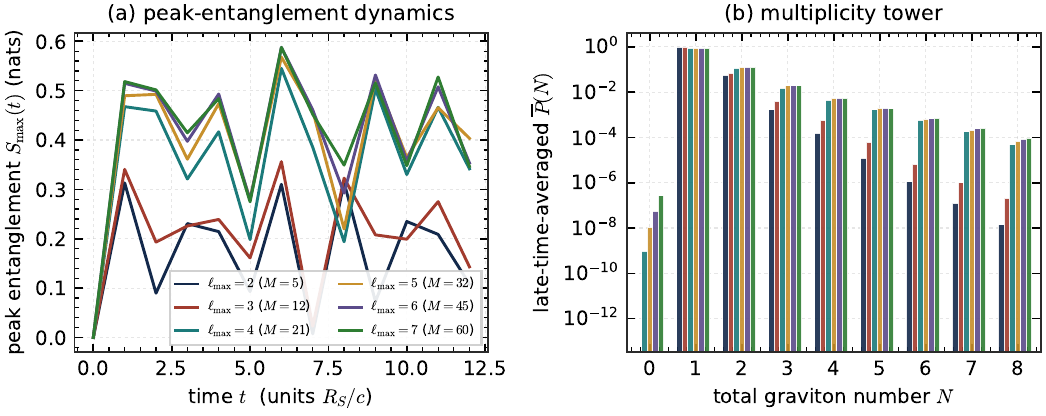}
\caption{\label{fig:H3}(a)~Peak mode-bipartition entanglement $S_{\max}(t)$
along the trajectory for each multipole depth: bounded and, for
$\ell_{\max}\ge4$, collapsing onto a common envelope. (b)~Late-time-averaged
multiplicity tower $\overline{P}(N)$, dominated by $N=1$ and falling steeply,
its multiparticle tail lengthening as the register grows.}
\end{figure}

\paragraph{Results.}
The single-$(2,0)$ quench was evolved to $t=12\,\Rs$ at $\gt=12$, $d=4$, for
$\ell_{\max}=2,\dots,7$; the late-time-averaged observables are collected in
Table~\ref{tab:climb}. Three of them saturate: the inelasticity at
$\bar\eta_\infty\simeq0.156$, the mean occupation at $\bar N\simeq1.20$, and
the peak mode-bipartition entanglement at $\overline{S}_{\max}\simeq0.45$ nats.
The saturated entanglement is far below the Page value for sixty modes, and its
late-time profile (Fig.~\ref{fig:H1}) peaks near the low-$|m|$ active modes and
decays across the chain, the $\ell_{\max}=4$--$7$ profiles collapsing onto a
common envelope: an area-law dressing whose structure has converged. The
multiplicity tower $\overline{P}(N)$ (Fig.~\ref{fig:H3}) is dominated by $N=1$
and falls steeply, its tail lengthening as the register grows, and the peak
entanglement oscillates within a bounded envelope throughout the trajectory
(Fig.~\ref{fig:H3}a). The mid-chain half-cut entropy is not a faithful
size-independent measure in this ordering: with $\pm m$ partners adjacent the
central bond falls in the weakly-populated high-$|m|$ tail, so $S_{1/2}$ is
position-dependent and non-monotonic in $M$, and the peak $\overline{S}_{\max}$
together with the full profile are the physical measures.

\paragraph{Convergence.}
Bond dimension $\chi=24$ is sufficient. At $\ell_{\max}=4$ the late-time
inelasticity is $\bar\eta=0.1360,0.1369,0.1371,0.1360$ at $\chi=24,32,48,96$,
a spread below $1\%$, and the half-cut entropy varies by about $4\%$ over the
same range: the observables reside in the largest Schmidt values and are
insensitive to the discarded tail. The final truncation error grows with depth
but stays below the Trotter-level $10^{-3}$ at every register
(Fig.~\ref{fig:H2}d). The per-mode occupation cutoff is
likewise converged, checked explicitly rather than by the mean-occupation
argument alone. Raising $d$ from $4$ to $8,16,32$ changes the late-time
inelasticity by at most $2.2\%$ (at $\ell_{\max}=4$), and $d=8$, $16$ and $32$
give identical results at $\ell_{\max}=2,3$, the occupation having saturated
(Table~\ref{tab:dconv}). The $d=8$ runs form a second family of large registers
on the occupation axis, reaching $96$ qubits at $\ell_{\max}=5$; the mode-axis
climb ($d=4$) and the occupation-axis refinement ($d=8$) agree across the tower
(Fig.~\ref{fig:H2}), so the saturated observables are robust to both
truncations.

\paragraph{Scope.}
The climb settles the size-limited dynamical and entanglement questions of
Sec.~\ref{app:scaling:physics}: the perturbative rigidity and the sub-Page
entanglement persist to $M=60$ modes, and the inelasticity converges to a
finite value rather than running away. It does not extend the interior
level-spacing statistics of Sec.~\ref{app:scaling:size}, which require many
highly entangled interior eigenstates in each fixed-$(L^2,L_z)$ sector, outside
the reach of the tensor-network evolution used here; that observable remains at
its exact-diagonalization registers.

\begin{table}[t]
\caption{\label{tab:climb}Late-time-averaged observables of the single-$(2,0)$
quench along the multipole ladder, from $L_z$-graded matrix-product-state
evolution ($\gt=12$, $d=4$, $t=12\,\Rs$, bond dimension $\chi=24$). $M$ is the
mode count, $\chi_{\rm MPO}$ the Hamiltonian operator bond dimension,
$\bar\eta$ the inelasticity, $\bar N$ the mean occupation,
$\overline{S}_{\max}$ the peak mode-bipartition entanglement, and $\epsilon$
the final truncation error. The inelasticity, mean number, and peak
entanglement saturate by $M\gtrsim32$.}
\begin{ruledtabular}
\begin{tabular}{ccccccc}
$\ell_{\max}$ & $M$ & $\chi_{\rm MPO}$ & $\bar\eta$ & $\bar N$
& $\overline{S}_{\max}$ & $\epsilon$ \\
\hline
$2$ & $5$  & $12$  & $0.059$ & $1.06$ & $0.18$ & $\sim\!10^{-14}$ \\
$3$ & $12$ & $41$  & $0.072$ & $1.08$ & $0.22$ & $2\times10^{-6}$ \\
$4$ & $21$ & $114$ & $0.136$ & $1.17$ & $0.40$ & $2\times10^{-4}$ \\
$5$ & $32$ & $228$ & $0.152$ & $1.19$ & $0.43$ & $4\times10^{-4}$ \\
$6$ & $45$ & $429$ & $0.154$ & $1.19$ & $0.44$ & $5\times10^{-4}$ \\
$7$ & $60$ & $711$ & $0.156$ & $1.20$ & $0.45$ & $6\times10^{-4}$ \\
\end{tabular}
\end{ruledtabular}
\end{table}

\begin{table}[t]
\caption{\label{tab:dconv}Cutoff convergence of the late-time-averaged
observables. Raising the per-mode occupation cutoff $d$ (Gray-encoded in
$\lceil\log_2 d\rceil$ qubits per mode) leaves the inelasticity $\bar\eta$, the
peak entanglement $\overline{S}_{\max}$, and the mean occupation $\bar N$
essentially unchanged; $d=8$, $16$ and $32$ coincide at $\ell_{\max}=2,3$. The
$d=8$ family reaches $96$ qubits at $\ell_{\max}=5$. Bond dimension $\chi=24$,
$\gt=12$, $t=12\,\Rs$.}
\begin{ruledtabular}
\begin{tabular}{cccccc}
$\ell_{\max}$ & $d$ & qubits & $\bar\eta$ & $\overline{S}_{\max}$ & $\bar N$ \\
\hline
$2$ & $4$  & $10$ & $0.0588$ & $0.179$ & $1.061$ \\
$2$ & $8$  & $15$ & $0.0587$ & $0.180$ & $1.061$ \\
$2$ & $16$ & $20$ & $0.0587$ & $0.180$ & $1.061$ \\
$2$ & $32$ & $25$ & $0.0587$ & $0.180$ & $1.061$ \\
$3$ & $4$  & $24$ & $0.0715$ & $0.218$ & $1.077$ \\
$3$ & $8$  & $36$ & $0.0713$ & $0.216$ & $1.078$ \\
$3$ & $16$ & $48$ & $0.0713$ & $0.216$ & $1.078$ \\
$4$ & $4$  & $42$ & $0.1360$ & $0.395$ & $1.168$ \\
$4$ & $8$  & $63$ & $0.1330$ & $0.366$ & $1.175$ \\
$5$ & $4$  & $64$ & $0.1515$ & $0.429$ & $1.191$ \\
$5$ & $8$  & $96$ & $0.1499$ & $0.411$ & $1.196$ \\
$6$ & $4$ & $90$ & $0.1538$ & $0.442$ & $1.194$ \\
$7$ & $4$ & $120$ & $0.1559$ & $0.447$ & $1.196$ \\
\end{tabular}
\end{ruledtabular}
\end{table}

\subsection{Physics accessible only to larger systems}
\label{app:scaling:physics}

The limitations of Sec.~\ref{app:scaling:size} all trace to the small effective
dimension and small sector sizes of the statevector-accessible registers. The
tensor-network climb of Appendix~\ref{app:scaling:climb} enlarges the register
classically and addresses these questions in part; we take them in turn, marking
what the climb now settles and what remains for hardware.

\emph{(i) Out-of-time-order growth, a Lyapunov exponent, and the chaos bound.}
With many eigenstates participating, the squared commutator can develop a
genuine exponential window $C(t)\sim\epsilon\,e^{2\lambda_L t}$ preceding
saturation, from which a Lyapunov rate $\lambda_L$ may be extracted and the
Maldacena--Shenker--Stanford bound $\lambda_L\le2\pi T$ tested \cite{MSS2016};
saturation of that bound is the signature of the maximal chaos expected of a
fast-scrambling horizon \cite{SekinoSusskind2008}. The measurement is
instrument-ready: the ancilla Hadamard-test pipeline reproduces the analytic
squared-commutator rate $2\Omega$ of the exactly solvable inverted-oscillator
benchmark to $0.2\%$, and the symmetry-exact total-occupation truncation was
constructed precisely to render out-of-time-order correlators at larger $\Nmax$
accessible. The resonance-free spectrum that
forbids near-resonant cubic processes is a property of the boost dispersion at
every multipole and does not weaken with system size, so the leading-soft
self-interaction may remain non-scrambling however large the register; the
rising level-spacing ratio points the other way. A larger simulation is
therefore a decisive test of which ingredient of the near-horizon dynamics, the
cubic self-interaction, the exchange/eikonal sector, or beyond-leading-soft
kernels, is responsible for the maximal chaos attributed to the horizon. The climb bears
on this: the dressing remains area-law and non-scrambling to sixty modes
(Appendix~\ref{app:scaling:climb}), consistent with the resonance-free spectrum,
so no exponential window appears up to that size. Extracting a Lyapunov rate,
however, requires a genuinely chaotic window; the definitive test needs either a regime in which the dynamics
scramble or the hardware execution the truncation was built for.

\emph{(ii) Eigenstate thermalization.} When the effective dimension is large the
energy shell contains many eigenstates rather than the single eigenstate of the
present registers, and the eigenstate-thermalization hypothesis becomes directly
testable \cite{Deutsch1991,Srednicki1994,DAlessio2016}: one can ask whether the
late-time multiplicity distribution approaches the microcanonical, and the
$\beta$-matched thermal, prediction, a genuine thermalization, in place of the
single dressed eigenstate ($d_{\mathrm{eff}}=1.06$) found here. This is the
direct probe of whether the near-horizon self-interaction thermalizes the
longitudinal sector, a question the single-eigenstate quench of the accessible
registers cannot even pose. The climb constrains this too: to sixty modes the
late-time multiplicity distribution remains far from the thermal and
microcanonical references and the mean occupation stays small
($\bar N\simeq1.20$), showing no approach to thermalization. A direct
eigenstate-resolved test nonetheless requires many interior eigenstates in a
shell, outside the reach of the ground-state-oriented tensor-network evolution,
and remains open.

\emph{(iii) Approach to the Page curve.} The mode-bipartition entanglement
measured at $\ell=2$ sits $2.3$ nats below the Haar (Page) value. The climb now
answers whether it climbs toward Page as the Hilbert space grows: it does not.
The peak entanglement saturates at $\simeq0.45$ nats through $M=60$ modes and its
profile is area-law (Appendix~\ref{app:scaling:climb}, Figs.~\ref{fig:H1}
and~\ref{fig:H2}), far below the Page value at every register. The leading-soft
cubic vertex therefore does not generate Page-like entanglement by itself;
whatever sources the Page curve in the near-horizon program, it is the
exchange/eikonal sector or beyond-leading-soft kernels, not this vertex
\cite{Page1993}. This is the sharpest form of the perturbative-rigidity result.

\emph{(iv) The multiparticle tower at large occupation.} Raising the per-mode
occupation cutoff to $d=8,16,32$ and climbing to $\ell_{\max}=7$ resolves the
tower directly. The late-time distribution $\overline{P}(N)$ is measured to
$M=60$ modes, and to $96$ qubits at larger occupation
(Appendix~\ref{app:scaling:climb}): it is dominated by $N=1$, falls steeply, and
is converged in the cutoff, the occupation having saturated by $d=8$. The tower
is bounded and non-thermal, with no sign of the broad $2\to2N$ distribution that
a peak multiplicity $\Nmax\sim(2/e)\Sbh$ would require; a compatible comparison
with the exchange-sector prediction thus favours distinct mechanisms for the
self-interaction and the exchange channel.

Taken together, the classical climb has confirmed the perturbative rigidity as a
genuine dynamical property of the leading-soft self-interaction across the full
multipole register: the entanglement stays area-law (iii) and the multiparticle
tower stays bounded and cutoff-converged (iv). The two questions it cannot
settle, the Lyapunov rate (i) and eigenstate thermalization (ii), are precisely
those that presuppose the dynamics to \emph{leave} the rigid regime, at stronger
coupling or through the exchange sector, and both need many highly entangled
interior states. That is the regime reserved for hardware: Table~\ref{tab:scaling}
fixes the register size and circuit budget at which the hardware tests become
accessible, and the crossover to volume-law entanglement, not the qubit count,
is what triggers them.

\section{Supplementary figures}
\label{app:supp}

\begin{figure}[H]
	\includegraphics[width=0.9\columnwidth]{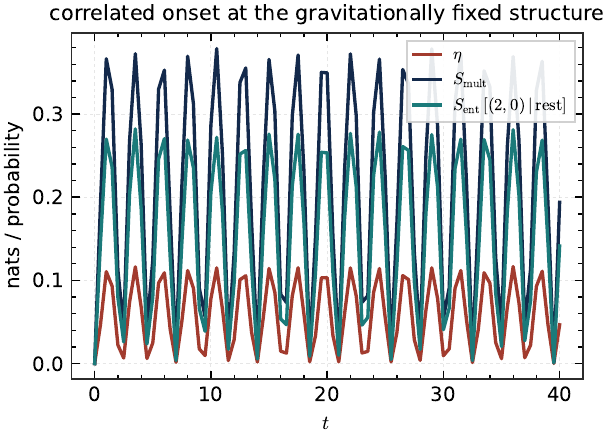}
	\caption{\label{fig:C3}Synchronized, oscillatory growth of $\eta$, $S_{\rm mult}$, and the single-mode entanglement entropy at $\gt=12$.}
\end{figure}

\begin{figure}[h]
	\includegraphics[width=0.9\columnwidth]{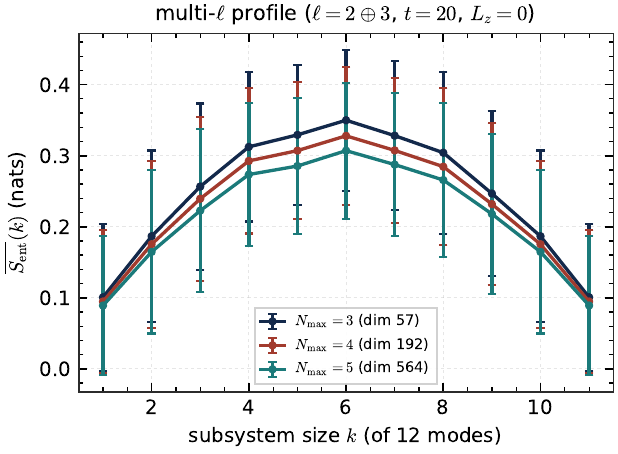}
	\caption{\label{fig:C4}Multi-$\ell$ entanglement profile under the su(2)-exact truncation, $\Nmax=3,4,5$: sub-Haar and converging.}
\end{figure}

\begin{figure}[H]
	\includegraphics[width=0.9\columnwidth]{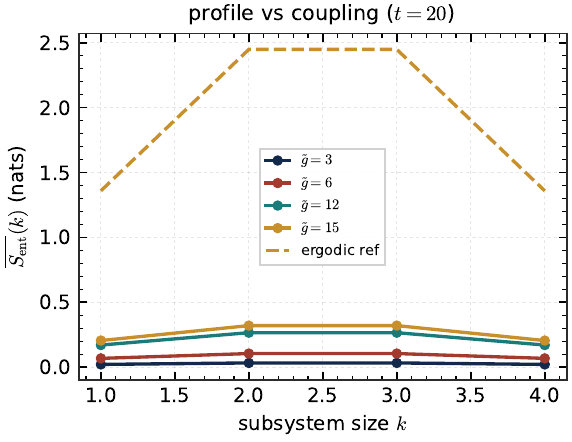}
	\caption{\label{fig:C2s}Entanglement profile versus coupling at $t=20$.}
\end{figure}

\begin{figure}[h!]
	\includegraphics[width=0.9\columnwidth]{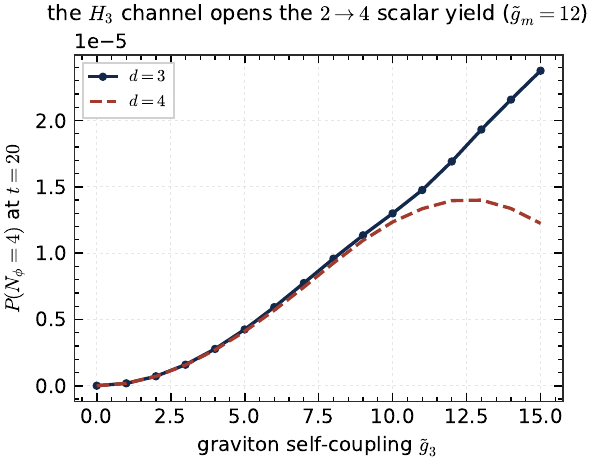}
	\caption{\label{fig:F2s}Four-scalar yield versus $\gtthree$ with the $d=4$
		truncation overlay ($21\%$ shift at $\gtthree=12$).}
\end{figure}

\end{document}